\pdfoutput=0
\documentclass[12pt,onecolumn,journal]{IEEEtran}
\usepackage{cite}
\usepackage{amsmath,amssymb,amsfonts}
\usepackage{graphicx,color}
\usepackage{comment}
\usepackage{graphicx}
\usepackage{textcomp}
\usepackage{psfig}
\usepackage{steinmetz} 
\usepackage{tabularx}
\usepackage[multiple]{footmisc} 
\usepackage{algorithm}
\usepackage{algpseudocode}

\newcommand{\mycomment}[1]{%
}%
\usepackage{environ}
\NewEnviron{nestedcomment}{\mycomment{\BODY}}

\DeclareMathOperator{\sinc}{sinc}

\def\BibTeX{{\rm B\kern-.05em{\sc i\kern-.025em b}\kern-.08em
    T\kern-.1667em\lower.7ex\hbox{E}\kern-.125emX}}
\begin{document}
\title{Achieving Optimum Received Power for Discrete-Phase RISs with Elementwise Updates in the Least Number of Steps
\thanks{This work is partially supported by NSF grant 2030029.}
}
\ifCLASSOPTIONonecolumn
\author{\IEEEauthorblockN{Dogan Kutay Pekcan, {\em Student Member, IEEE}}\\
\IEEEauthorblockA{\textrm{CPCC, Department of EECS} \\
\textrm{University of California, Irvine, CA, USA}\\
dpekcan@uci.edu\\[7mm]}
\and
\IEEEauthorblockN{ Ender Ayanoglu, {\em Fellow, IEEE}}\\
\IEEEauthorblockA{\textrm{CPCC, Department of EECS} \\
\textrm{University of California, Irvine, CA, USA}
\\
ayanoglu@uci.edu
}}
\else
\author{\IEEEauthorblockN{Dogan Kutay Pekcan, {\em Student Member, IEEE}}
\IEEEauthorblockA{\textrm{CPCC, Department of EECS} \\
\textrm{University of California, Irvine, CA, USA}\\
dpekcan@uci.edu\\}
\and
\IEEEauthorblockN{Ender Ayanoglu, {\em Fellow, IEEE}}
\IEEEauthorblockA{\textrm{CPCC, Department of EECS} \\
\textrm{University of California, Irvine, CA, USA}\\
ayanoglu@uci.edu
}
}
\fi

\maketitle

\begin{abstract}
The problem of optimizing discrete phases in a reconfigurable intelligent surface (RIS) to maximize
the received power at a user equipment is addressed.
Necessary and sufficient conditions to achieve this maximization are given.
These conditions are employed in an algorithm to achieve the maximization. New versions of the algorithm are given
that are proven to achieve convergence in $N$ or fewer steps whether the direct link is completely
blocked or not, where $N$ is the number of the RIS elements, whereas previously published results achieve this in $KN$ or $2N$ number
of steps where $K$ is the number of discrete phases. Thus, for a discrete-phase RIS,
the techniques presented in this paper achieve the optimum received power in the smallest number of steps
published in the literature. In addition, in each of those $N$ steps, the techniques presented in this paper
determine only one or a small number of phase shifts with a simple elementwise update rule, which result in
a substantial reduction of computation time, as compared to the algorithms in the literature. As a secondary result, we define the uniform polar quantization (UPQ) algorithm which is an intuitive quantization algorithm that can approximate the continuous solution with an approximation ratio of $\sinc^2(1/K)$ and achieve low time-complexity, given perfect knowledge of the channel.
\end{abstract}


\section{INTRODUCTION}\label{ch:0}
%
\IEEEPARstart{I}{n} wireless communications, the performance demand has significantly grown with the advancements in technology, requiring to cope with many challenges in the propagation environment. In dense urban outdoor-indoor areas, the problems of shadowing and fading can become severe. In specific scenarios where the direct link between the base stations and users is blocked, despite the more extensive beamforming capability of the base stations, the performance can be significantly degraded \cite{ABCHLSZ14}. In order to catch up with the ever-growing quality of service (QoS) and energy efficiency requirements in mobile communications, the challenges due to blockages must be overcome to prevent performance degradation \cite{BDDDAZ19}.

A reconfigurable intelligent surface (RIS), also known as intelligent reflective surface (IRS), can manipulate the incident electromagnetic waves to control the propagation environment by varying the phases of the incident signals with its low-cost passive reflecting elements \cite{ACS22}. In recent years, RISs have been studied in many communication systems in the literature, and it is shown that RIS can become a crucial enabler for communication environments, especially when there is a loss in line-of-sight (LOS) between the base station (BS) and user equipment (UE) \cite{BWMPSC22},\cite{PYTCLWZB21}.

For complex communication scenarios, algorithms are developed with the assumption of continuous phase shifts at the RIS, for ease of optimization \cite{YXS20},\cite{YXS19},\cite{WZ19},\cite{QRLKA21}. In \cite{YXS20}, to achieve a globally optimal solution, the authors developed an algorithm which is based on the branch-and-bound method. In \cite{YXS19}, the authors developed a secure wireless communication system for a single-user case by adopting the majorization-minimization and block coordinate descent (BCD) techniques. Assuming perfect knowledge of the channels in \cite{WZ19}, the authors jointly designed the active and passive beamforming to minimize the transmit power, subject to individual signal-to-interference-ratio constraints of the users. In \cite{QRLKA21}, the authors consider an RIS-aided multi-antenna transmission with the effect of channel fading and phase noise impairments at the RIS, to maximize the signal-to-noise-ratio (SNR) at the receiver.

Practical large RISs can usually employ quantized phase shift as they are more cost-effective \cite{WZ20}. A two-stage approach to address the discrete phase shifts constraints is to project the continuous solution to the closest value in the discrete set \cite{WZ20,YBA20},\cite{YZZ20,ZSRLCL22}. Although, as shown in the present paper, the quantization approach can potentially provide high performance and low computational complexity, it can only give marginal insight on the actual optimum discrete phase shifts selection problem.

When it comes to designing algorithms with discrete phase shift constraints, as the number of possible solutions increases exponentially with the number of RIS elements, exponential search techniques are required \cite{XDMWQ22}, and a closed-form solution is practically unavailable, as we discuss in Section~IV-A.
As an example, the authors in \cite{b1} stated that the discrete beamforming problem for the RIS turns out to be a generally NP-hard discrete quadratic program (QP), and a generic $K$-ary discrete QP remains as an open problem. To guarantee the optimal solution, most of the prior work in single-user scenarios had exponential complexity, sometimes over the number of RIS elements ($N$) \cite{WZ20}, and sometimes over the phase shift levels ($K$) \cite{JSDH21}. In this regard, probabilistic optimization techniques have also drawn attention \cite{LPW01},\cite{YKDO13,PD23}. In \cite{LPW01}, the idea of probabilistic data association (PDA) is used to address general binary quadratic problems (BQPs) for multi-user detection in code division multiple access. Machine-to-machine wireless communication with high reliability is considered in \cite{YKDO13}, where authors approached BQPs by developing a PDA algorithm, achieving near-optimal results. Recently, authors in \cite{PD23} developed a comprehensive probabilistic technique to address the quantization error and scalability issues in a variety of discrete RIS optimization problems, where the proposed technique outperforms general approaches such as the closest point projection (CPP) method given in \cite{ZSRLCL22}.

To the best of our knowledge, there is still a gap in the literature for further research in the optimal discrete beamforming problem. Regarding the previously published work addressing the optimal discrete beamforming, the authors in \cite{SBRFZT21} pointed out that an $N$-step search algorithm for $K$-ary beamforming can be developed. In \cite{ZSRLCL22}, the authors provided 
an optimal algorithm for the binary case, where $2N+2$ steps are required for convergence. Also in \cite{ZSRLCL22}, the authors provided an approximation algorithm (APX), and a simple quantization algorithm for $K$-ary beamforming, where the performance is evaluated over the estimated channels. The authors in \cite{9961233} proposed a $K$-ary optimal discrete beamforming algorithm with a polynomial search complexity, i.e., $2N(K-1)$ steps to ensure optimality. Recently, the authors in \cite{b1} proposed the first $K$-ary linear time algorithm with $KN$ steps to converge, which could be reduced to $2N$ steps when the direct link is not completely blocked.

In this work, we address the problem of optimizing discrete phases in an RIS to maximize the received power at a UE for a single-input single-output (SISO) system with full channel state information (CSI). We also extend our solutions to special cases of multiple-input multiple-output (MISO), multiple users, and imperfect CSI scenarios. The main contributions are given as follows:
\begin{itemize}
	\item We provide necessary and sufficient conditions to achieve this maximization. We employ these conditions to develop a linear time algorithm achieving provable optimality convergence in $N$ or \textit{fewer} steps, whether the direct link between the BS and the UE is blocked or not. Employing the geometric approach in \cite{b1}, we develop a formal elementwise update rule to be used, so that in each of those $N$ or \textit{fewer} steps, one or a small number of elements are updated, respectively.
	\item  We prove the periodicity in elementwise updates for the optimal beamforming, and exploit this periodicity to both reduce the number of steps, and to provide a simple initialization for the search algorithms. With this, we show the relation between the channel phases and the pattern of elements in which the periodicity must occur.
	\item Our developed algorithms are shown to give a substantial reduction of computation time, as compared to the algorithms in the literature \cite{ZSRLCL22}, \cite{XDMWQ22}, \cite{b1}, \cite{SBRFZT21}. Besides providing improved optimal discrete phase shift selection algorithms, this work gives further insights on the discrete beamforming optimization, with the periodicity rule.
	\item  We formally define an intuitive, quantization-based, discrete phase shift selection algorithm, which we will call uniform polar quantization (UPQ). We show that provided full CSI is available, it performs well, in terms of both performance and the computational time complexity. With UPQ, we reveal that the optimal discrete beamforming can be closely approximated as a quantization solution.
\end{itemize}
The received power maximization problem in this work is connected to generic $K$-ary discrete QP, where the objective is to maximize $\mathbf{x}^{\mathrm{T}}\mathbf{Q}\mathbf{x}$, with $\mathrm{rank}(\mathbf{Q}) = 1$ \cite{b1,XDMWQ22}, where the single eigenvector is positive. Therefore, the linear time algorithms with the elemental update rules in the present paper can be employed in similar maximization problems, so that the global optimum can be achieved in linear time.

\section{SYSTEM MODEL} \label{sec:signalmodel}
Consider a point-to-point communication scenario aided by an RIS, where there is a non-line-of-sight (NLOS) channel between the BS and the UE. The RIS has $N$ elements located over $N_z$ rows and $N_y$ columns, in a uniform planar array (UPA) structure, as shown in Fig. \ref{fig:ris}.
We consider $K$ discrete phase shifts for the RIS, i.e., $\theta_n \in \Phi_K$ and $\Phi_K = \{\omega, 2\omega, \ldots, K\omega\}$ with $\omega=\frac{2\pi}{K}$ and $j=\sqrt{-1}$. The set $\Phi_K$ can equivalently be described as $\{ 0, \omega, 2\omega, \ldots , (K-1)\omega\}$. Hence, the RIS introduces amplitude $\beta^R_n$ and phase shift $\theta_n$ for $n = {1, 2, \dots, N}$ in the $N$-element reflection coefficient vector
\begin{equation}
    {\bf w} = \left[ \beta^R_1 e^{j\theta_1}, \beta^R_2 e^{j\theta_2} \dots, \beta^R_N e^{j\theta_N} \right]
\end{equation}
where for practicality, we let $\beta^R_n = 1$ in this paper. Let $s \in \mathbb{C}$ be the transmitted symbol. The received signal at the UE is given as \cite{WZ19}
\begin{equation} \label{eq:rxSISO}
    y= ({\bf h}_u^H {\bf W} {\bf h}_b + h_0) s+z,
\end{equation}
where ${\bf W} = {\rm diag}({\bf w})$, $h_0 \in \mathbb{C}$ is the direct link between the BS and UE, ${\bf h}_u \in \mathbb{C}^{N \times 1}$ and ${\bf h}_b \in \mathbb{C}^{N \times 1}$ are the equivalent channels of the RIS-UE and BS-RIS links, respectively, and $z$ is the additive white Gaussian noise (AWGN) at the UE antenna.

Let ${\bf h} = {\bf h}^*_u \odot {\bf h}_b$, where $\odot$ is the elementwise (Hadamard) multiplication of the two vectors, and the additive noise $z$ be a complex Gaussian random variable with variance $\sigma^2$, i.e., $z \sim \mathcal{CN}(0,\sigma^2)$. Assuming a mean power constraint at the BS, i.e., $\mathbb{E}[|s|^2] \leq P$, the achievable ergodic data rate in bps/Hz is given by
\begin{equation} \label{eq:ergodicRate}
    \gamma = \mathbb{E}\left[ \log \left(1 + \frac{P}{\sigma^2}\left| h_0 + {\bf h}^T {\bf w} \right|^2 \right) \right].
\end{equation}
Therefore, the maximization of the channel power inside the logarithm in (\ref{eq:ergodicRate}) amounts to maximizing the ergodic data rate, which is a commonly used performance metric in wireless communications. With this, the problem of maximizing the overall channel gain, i.e., the received signal power by performing discrete beamforming at the RIS is given as
\begin{equation}\label{eq:P1}
    {\rm (P1)} \max_{\phase{ [{\bf w}]_n }\in \Phi_K} |h_0 + {\bf h}^T {\bf w}|^2.
\end{equation}
The received power maximization problem in (P1) is shown to belong to the class of general quadratic programming problems, i.e., $K$-ary discrete QP, where the objective is to maximize $\mathbf{x}^{\mathrm{T}}\mathbf{Q}\mathbf{x}$, with $\mathrm{rank}(\mathbf{Q}) = 1$, which has been known to be NP-hard \cite{ZH06,b1,XDMWQ22}. Furthermore, authors in \cite{XDMWQ22} show that the problem (P1) can be reformulated as
\begin{equation}
    {\rm (P2)} \max_{\phase{ \bar{w}_1},\dots,\phase{ \bar{w}_{N+1}} \in \Phi_K} |{\bf \Tilde{g}}^H{\bf \Bar{w}}|^2
\end{equation}
where ${\bf \Bar{w}} = [e^{j\theta_1}, e^{j\theta_2}, \dots, e^{j\theta_{N}}, e^{j\theta_{N+1}}]^T$ and $\Tilde{{\bf g}}$ corresponds to the unique eigenvector of ${\bf \Tilde{Q}}$ with ${\bf \Tilde{Q}} = [{\bf h}, h_0][{\bf h}, h_0]^H$. So, the solution of (P1) can be extracted as ${\bf \hat{w}} = \frac{{\bf \bar{w}}(1:N)}{{\bf \bar{w}}(N+1)}$, where ${\bf \bar{w}}(1:N)$ correspond to the first $N$ elements in ${\bf \bar{w}}$ and ${\bf \bar{w}}(N+1)$ is the element at the last index. Therefore, the problem of interest is the maximization of the inner product either in (P1) or (P2) with discrete phase shift constraints at the RIS\footnote{Strictly speaking, it is the magnitude squared of a complex-valued inner product, but we will use the informal statement ``inner product'' as in \cite{XDMWQ22}.}.

The equivalent maximization problems (P1) and (P2) frequently arise in RIS-aided communications systems.
For example, in RIS-assisted single-user multiple-input multiple-output (MIMO) localization systems, in two-stage approaches to refine the localization performance, the objective boils down to solving the discrete beamforming design for the RIS to maximize the passive beamforming gain \cite{LDDHCHB23,HWSSJ20}.
In fact, in \cite{LDDHCHB23}, the authors develop a discrete beamforming approach named Fast Passive Beamforming (FPB) algorithm that can achieve the global optimum for the very same problem in (P2) and used for improving the localization performance in mmWave MIMO.
Also, a joint framework for channel estimation and passive beamforming for an RIS is developed with the discrete phase shift constraints at the RIS \cite{YZZ20}, where the exact problem (P2) needs to be solved for the channel-gain-maximization approach in the progressive passive beamforming refinement.

The special case of single-antenna assumption has importance and draws attention in the literature, as problems (P1) and (P2), especially for analyzing the potential of the RIS in general scenarios \cite{BWMPSC22}, practical codebook design and optimization for RISs \cite{ACME23,GJSCES23}, and experimental setups with an RIS where the system is tested and optimized with discrete phase shifts \cite{PYTCLWZB21,TTKSMOKSCA22,KTCT20}. Therefore, having an efficient algorithm with provable global optimality as a benchmark is crucial to properly assess the performance of the developed algorithms in experimental setups, as well as to determine how reliable the direct quantization approach can be under different channels.

Beyond the extent of (P1) and (P2), ample research in the context of MISO and MIMO systems with discrete beamforming at the RISs has been conducted in the recent literature \cite{HZZZZH23,WGWM23}. In \cite{HZZZZH23}, authors performed joint active and passive beamforming with the generalized Benders decomposition (GBD)-based algorithm with codebook-based passive beamforming at multiple RISs for a multi-user MISO system to minimize the transmit power subject to signal-to-interference-and-noise-ratio (SINR). The results are shown to approximate the global optimum. In \cite{WGWM23}, the authors consider a RIS-aided MIMO system with transceiver hardware impairments, where the aim is to minimize the total mean squared errors (MSEs) of multiple data streams. They propose a two-tier majorization-minimization (MM) based and a modified Riemannian gradient descent (RGA) algorithm to obtain the sub-optimal solution of the RIS reflection matrix. Thus, regarding the recent work with more comprehensive scenarios and problems than (P1) and (P2), the aim of maximizing the channel gain has not attracted attention due to the non-convex and NP-hard constraints as well as the solutions can only approximate the global optimum solution.

\subsection{Special Case of Achieving Global Optimum with MISO}

In this section, we remark on a special case of a point-to-point downlink MISO communication scenario aided by an RIS, by considering a special case similar to \cite{WGWM23} to achieve optimality, where the direct BS-UE link is blocked. With this, (\ref{eq:rxSISO}) can be rewritten as
\begin{equation} \label{eq:rxMISO}
    y= {\bf h}_u^H {\bf W} {\bf G} {\bf x} s+z,
\end{equation}
where ${\bf x} \in \mathbb{C}^M$ is the transmit beamforming vector, ${\bf G} \in \mathbb{C}^{N \times M}$ is the equivalent channel of the BS-RIS link, and ${\bf W} = {\rm diag}({\bf w})$. The channel from RIS to UE is denoted by ${\bf h}_u \in \mathbb{C}^N$.
Taking the far-field regime into account with a physical model that is based on angle directions \cite{MZXWZQ23}, the received signal power in equation (\ref{eq:rxMISO}), i.e., $|y|^2$, can be expressed as
\begin{equation}\label{eq:rxMISOangle}
    |y|^2 = |{\bf bWCx}|^2
\end{equation}
where ${\bf C} = [{\bf a}(\vartheta_{b,1},\varphi_{b,1}), {\bf a}(\vartheta_{b,2},\varphi_{b,2}), \cdots, {\bf a}(\vartheta_{b,M},\varphi_{b,M})]$,
$\mathbf{b}=\mathbf{a}^H(\vartheta_u,\varphi_u)$,
with $\{\vartheta_u, \varphi_u\}$ is the departing direction of the reflected signal
and $\{\vartheta_{b,m},\varphi_{b,m}\}, m=1,\dots,M$ is the arriving signal direction from BS antenna $m$. Here, ${\bf a}(\vartheta,\varphi)$ is the array response vector defined in (\ref{eq:arrayresp}) in the next section.

Similar to the analysis in the Appendix of the fifth version of \cite{XDMWQ22}, equation (\ref{eq:rxMISOangle}) can be rewritten as
\begin{align}
    |y|^2 &= |{\bf bWCx}|^2 \nonumber \\
    &= {\bf w}^H ({\bf P}^T \odot {\bf R}) {\bf w} \nonumber \\
    &= {\bf w}^H {\bf Q} {\bf w}, \label{eq:MISOpower}
\end{align}
where ${\bf Q} = {\bf P}^T \odot {\bf R}$, ${\bf P} = {\bf Cx} {\bf x}^H {\bf C}^H$, and ${\bf R} = {\bf b}^H {\bf b}$. Since ${\bf Q}$ is a semi-positive definite matrix with rank one, maximizing (\ref{eq:MISOpower}) boils down to maximizing $|{\bf z}^H {\bf w}|$, where ${\bf z}$ is the eigenvector associated with the maximum eigenvalue of ${\bf Q}$.

In this paper, one of the primary objectives is to address the problem in (P1), equivalently (P2), with a highly efficient algorithm. In fact, to the best of our knowledge, our proposed algorithms are the fastest converging to the global optimum in the literature to perform discrete beamforming optimization for (P1), and a general $K$-ary discrete quadratic programming. The extension of this framework {\it with elementwise updates} for a general MISO scenario is left as a future work. In Section~IV, we define the problem formally and then introduce several algorithms to solve it. 
\section{CHANNEL MODEL} \label{sec:channelmodel}
\begin{figure}[!t]
\centering
\includegraphics[width=0.55\textwidth]{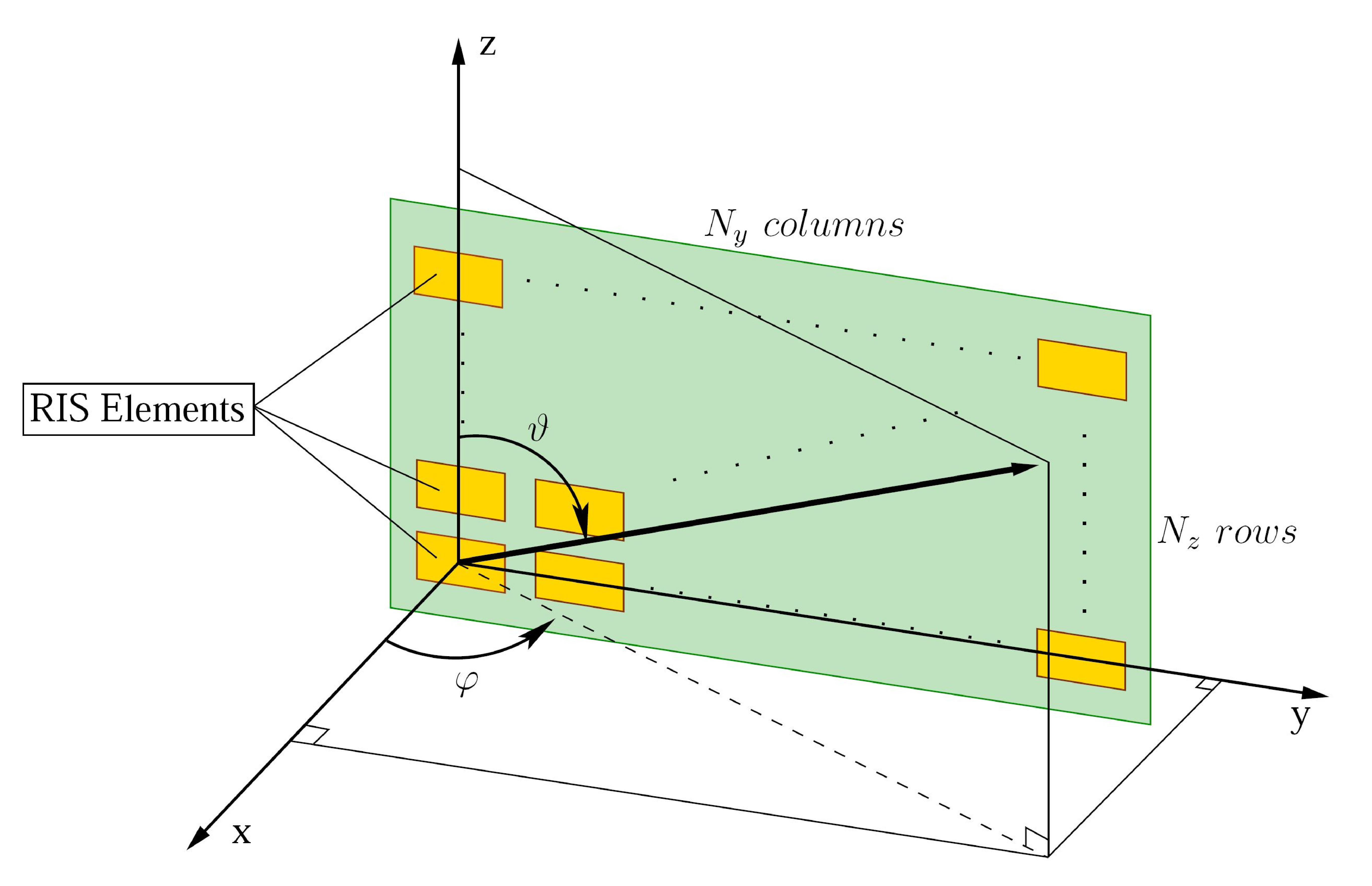}
\caption{RIS structure.}
\label{fig:ris}
\end{figure}

This section describes the channel model we employed. The RIS is placed such that the origin is at the first row and the first column of the RIS, as shown in Fig. \ref{fig:ris}. We let $\{\vartheta_b, \varphi_b\}$ and $\{\vartheta_u, \varphi_u\}$ pairs be the elevation and azimuth angles for the BS and the UE. The array response vector for the RIS is calculated by considering two uniform linear arrays (ULA) along the $y$-axis and the $z$-axis with array response vectors ${\bf a}_{y}(\vartheta, \varphi)$ and ${\bf a}_{z}(\vartheta)$, which are calculated as
\begin{align*}
        &\mathbf{a}_{y} =\left[1, e^{-j 2 \pi \frac{d_{y}}{\lambda} \sin\vartheta \sin\varphi},\cdots, e^{-j 2 \pi (N_y-1) \frac{d_{y}}{\lambda} \sin\vartheta \sin\varphi}\right]^{\mathrm{T}}\\
        &\mathbf{a}_{z}(\vartheta) =\left[1, e^{-j 2 \pi \frac{d_{z}}{\lambda} \cos\vartheta}, \cdots, e^{-j 2 \pi (N_z-1) \frac{d_{z}}{\lambda} \cos\vartheta}\right]^{\mathrm{T}}
\end{align*}
where $d_{y} \leq \lambda / 2$ and $d_{z} \leq \lambda / 2$ and where we used the shorthand notation $\mathbf{a}_{y}$ for $\mathbf{a}_{y}(\vartheta,\varphi)$. In this formulation, $d_{y}$ and $d_{z}$ are the spacing between the elements in respective axes, and $\lambda$ is the wavelength of the incident signal. Therefore, for the described uniform planar array (UPA) structure, the array response vector can be calculated as
\begin{equation} \label{eq:arrayresp}
    \mathbf{a}(\vartheta, \varphi)=\mathbf{a}_{y}(\vartheta, \varphi) \otimes \mathbf{a}_{z}(\vartheta)
\end{equation}
where $\otimes$ is the Kronecker product and $\mathbf{a}(\vartheta, \varphi) \in \mathbb{C}^{N}$. Note that, in $\mathbf{a}(\vartheta, \varphi)$, elements are ordered column-wise.

We employ the channel model from \cite{JCY21,WZ20}, where the BS-RIS link ${\bf h}_b$ undergoes Rician fading:
\begin{equation}\label{eq:RicianFading}
    {\bf h}_b = {\rm PL}_b \left( \sqrt{\frac{\kappa}{1+\kappa}} {\bf h}^{\rm LOS}_b + \sqrt{\frac{1}{1+\kappa}} {\bf h}^{\rm NLOS}_b \right)
\end{equation}
where ${\rm PL}_b = 30+22 \log_{10}(d_b)$ is the path loss (in dB) in the BS-RIS link with $d_b$ being the distance between the BS and RIS, the LOS component is calculated with the RIS steering vectors given the angle-of-arrival (AoA) information of the BS, i.e., $\{\vartheta_b,\varphi_b\}$, with equation (\ref{eq:arrayresp}) as ${\bf h}^{\rm LOS}_b = \mathbf{a}(\vartheta_b, \varphi_b)$. The NLOS component, ${\bf h}^{\rm NLOS}_b \sim \mathcal{CN}({\bf 0},{\bf I}_{N\times N})$, consists of circularly symmetric complex Gaussian random variables and $\kappa$ is the Rician factor. The RIS-UE link ${\bf h}_u$ also undergoes Rician fading in (\ref{eq:RicianFading}) with the same parameters except for the distance and the AoA information, i.e., $d_u$ and $\{\vartheta_u,\varphi_u\}$, respectively.

The developed algorithms do not require the direct BS-UE link to be present to converge to the global optimum in the least number of steps. Whenever the direct link is not completely blocked, it follows Rayleigh fading, i.e., $h_0 = 10^{- \frac{{\rm PL}_0}{20}} \times \rho_0$, where, ${PL}_0(d_0) = 32.6+36.7 \log_{10}(d_0)$ is the path loss (in dB) with $d_0$ being the BS-UE distance and $\rho_0 \sim \mathcal{CN}(0,1)$ is a circularly symmetric complex Gaussian random variable.

To compare the performance and time complexity of our proposed algorithms with the existing methods from the literature with discrete phase shift constraints for the RIS, similar to \cite{b1}, we consider NLOS transmission in BS-RIS and RIS-UE channels with $\kappa=0$. The transmit power is 30 dBm with -90 dBm background noise power. Three-dimensional coordinate vectors $(-2,-1,0)$, $(50,-200,20)$, and $(0,0,0)$ are used as the locations of the RIS, BS, and the UE, respectively.

Finally, we remark that all of the developed algorithms can work not only with different Rician factors but also with arbitrary $\alpha_n$ selections.
Therefore, the proposed algorithms can be applied to a general $K$-ary discrete Quadratic Program, where the objective is to maximize ${\bf x}^T{\bf Q}{\bf x}$, with ${\rm rank}({\bf Q}) = 1$,
or equivalently, the inner product maximization of $|{\bf b}^T{\bf x}|^2$, as in (P2). 
\section{PROBLEM DEFINITION}\label{ch:1}
%
In this paper, we address
the problem of finding the values $\theta_1,
\theta_2, \ldots, \theta_N$ to maximize $| h_0 + \sum_{n=1}^N h_n e^{j\theta_n} |$ where $\theta_n
\in \Phi_K$ and $\Phi_K = \{\omega, 2\omega, \ldots, K\omega\}$ with $\omega=\frac{2\pi}{K}$ and
$j=\sqrt{-1}$. The
set $\Phi_K$ can equivalently be described as $\{ 0, \omega, 2\omega, \ldots , (K-1)\omega\}$.
The values $h_n\in \mathbb{C}$, $n=1,2,\ldots,N$ are the channel coefficients and
$\theta_n$ are the phase values added to the corresponding $h_n$ by a reconfigurable intelligent surface (RIS).

\begin{nestedcomment}
Towards achieving its goal, \cite{b1} introduced the following lemma.

{\em Lemma 1:\/} For an optimal solution $(\theta_1^*, \ldots, \theta_n^*)$ to problem (8),
each $\theta_n^*$ must satisfy
\begin{equation}\tag{11}
\theta_n^* = \arg \min_{\theta_n\in \Phi_K} |(\theta_n +\alpha_n - \phase{\mu}) \;{\rm mod}\; 2\pi|
\end{equation}
where $\phase{\mu}$ stands for the phase of $\mu$ in (10)\footnote{To prevent confusion,
we will use the same equation numbers (7)--(13) in \cite{b1}. Our own equation numbers,
not available in \cite{b1}, will begin at (19) and will be incremented from that number
on. Similarly, we will introduce Lemma~2 and Algorithm~2 in lieu of Lemma~1 and Algorithm~1
in \cite{b1}. Note that a lemma or an algorithm with
number 2 does not exist in \cite{b1}.}\footnote{In this paper, we define the ${\rm mod}$
function (the modulus function or the modulo operation) $x\; {\rm mod}\;y$ as the remainder after the
dividend $x>0$ is divided by the divisor $y>0$. We write it as $x\;{\rm mod}\; y$, $x\; (mod\; y)$, or
${\rm mod}\; (x, y)$. For $x<0$ and $y>0,$ we use the convention that the remainder should always
be the smallest such nonnegative number.}.

In \cite{b1}, problem (8) is defined as
\end{nestedcomment}
The problem can be formally described as
\begin{equation}
\begin{aligned}
 & \underset{\mbox{\boldmath$\theta$}}{\rm maximize\ } f({\mbox{\boldmath$\theta$}})\\
 & {\rm subject\ to\ } \theta_n\in \Phi_K,\ n=1, 2, \ldots, N
\end{aligned}
\label{eqn:eqn1}
\end{equation}
where
\begin{equation}
f({\mbox{\boldmath$\theta$}}) = \frac{1}{\beta_0^2}\bigg|\beta_0e^{j\alpha_0}+\sum_{n=1}^N \beta_n
e^{j(\alpha_n + \theta_n)}\bigg|^2,
\end{equation}
$h_n = \beta_ne^{j\alpha_n}$ for $n = 0, 1, \ldots, N$,
and ${\mbox{\boldmath$\theta$}} = (\theta_1, \theta_2, \ldots, \theta_N)$. Also, $g$ is defined as
\begin{equation}
g = h_0 + \sum_{n=1}^N h_n e^{j\theta_n^*}\label{eqn:eqn3}
\end{equation} where $\theta^*_n$ are the outcomes of the optimization in (\ref{eqn:eqn1}) and $\mu$ is defined as
\begin{equation}
\mu = \frac{g}{|g|}.\label{eqn:eqn4}
\end{equation}
Note that,
in (\ref{eqn:eqn1}), $\beta_0^2$ is a constant and therefore the maximization affects only the numerator.
For that reason, the case with $h_0=0$ can be taken care of by maximizing the numerator only.

The solution to the problem can be achieved by making use of the following lemma.

{\em Lemma:\/} For an optimal solution $(\theta_1^*, \theta_2^*, \ldots,
\theta_N^*)$, it is necessary and sufficient that each $\theta_n^*$ satisfy
\begin{equation}
\theta_n^* = \arg \max_{\theta_n\in \Phi_K} \cos(\theta_n + \alpha_n -\phase{\mu})
\label{eqn:lemma2}
\end{equation}
where $\phase{\mu}$ stands for the phase of
$\mu$ in (\ref{eqn:eqn4}).

{\em Proof:\/} We can rewrite (\ref{eqn:eqn3}) as
\begin{align}
|g| =& \ \beta_0 e^{j(\alpha_0-\phase{\mu})} + \sum_{n=1}^N \beta_n e^{j(\alpha_n+\theta_n^*-\phase{\mu})} \\
 = & \ \beta_0 \cos(\alpha_0 - \phase{\mu}) + j \beta_0 \sin (\alpha_0-\phase{\mu}) \nonumber\\
& + \sum_{n=1}^N \beta_n \cos(\theta_n^* + \alpha_n - \phase{\mu}) \nonumber\\
& + j \sum_{n=1}^N \beta_n \sin(\theta_n^* + \alpha_n - \phase{\mu}).
\label{eqn:absg}
\end{align}
Because $|g|$ is real-valued, the second and fourth terms in (\ref{eqn:absg}) sum to zero, and
\begin{equation}
|g| = \beta_0 \cos(\alpha_0 - \phase{\mu}) + \sum_{n=1}^N \beta_n \cos(\theta_n^* + \alpha_n - \phase{\mu}),
\end{equation}
from which (\ref{eqn:lemma2}) follows as a necessary and sufficient condition for the lemma to hold.
\hfill$\blacksquare$

Reference \cite{b1} attempts to decide a range of $\mu$ for which $\theta_n^* = k\omega$ must hold.
Towards that end, it first defines a sequence of complex numbers with respect to
each $n=1,2,\ldots,N$ as
\begin{equation}
s_{nk} = e^{j(\alpha_n+(k-0.5)\omega)},\ {\rm for}\ k=1,2,\ldots,K.
\label{eqn:eqn12}
\end{equation}
Then, \cite{b1} defines, for any two points $a$ and $b$ on the unit circle $C$, ${\rm arc}(a:b)$
to be the unit circular arc with $a$ as the initial end and $b$ as the terminal end in the counterclockwise
direction; in particular, it defines ${\rm arc}(a:b)$ as an open arc with the two endpoints $a$ and $b$
excluded. With this definition, \cite{b1} states the following proposition holds.

{\em Proposition 1:\/} A sufficient condition for $\theta_n^*=k\omega$ is
\begin{equation}
\mu \in {\rm arc} (s_{nk}:s_{n,k+1}).
\label{eqn:eqn8}
\end{equation}
Proposition~1 is compatible with the lemma given above. To see this, assume
$\mu$ satisfies (\ref{eqn:eqn8}). Then,
\begin{equation}
\phase{\mu}\in \Big(\alpha_n+\Big(k-\frac{1}{2}\Big)\omega,\alpha_n+\Big(k+\frac{1}{2}\Big)\omega\Big).
\end{equation}
Since $\omega=\frac{2\pi}{K}$,
\begin{equation}
\alpha_n-\phase{\mu}\in \Big((-2k-1)\frac{\pi}{K}, (-2k+1)\frac{\pi}{K}\Big)
\label{eqn:suffRange}
\end{equation}
considering the reversal of order due to the subtraction of $\phase{\mu}$.
Now, let $\theta_n = k\omega = 2k \frac{\pi}{K}$. Then
\begin{equation}
\theta_n + \alpha_n - \phase{\mu} \in \Big(-\frac{\pi}{K}, \frac{\pi}{K}\Big)
\end{equation}
and thus $\cos(\theta_n + \alpha_n-\phase{\mu})$ is the largest among
all other possibilities for $\theta_n$ because the slice $(-\frac{\pi}{K}, \frac{\pi}{K})$ corresponds to the largest values of the cosine function among all slices corresponding to different values of $\theta_n \in \Phi_K$ for $n=1,2,\ldots,N$.

We remark that, as an open arc, ${\rm arc}(a:b)$ does not contain the two endpoints $a$ and $b$. This implies that we omit the situation where $\mu = s_{nk}$ for some $n,k$, or similarly when $\mu$ lies right in between ${\rm arc} (s_{n,k-1}:s_{nk})$ and ${\rm arc} (s_{nk}:s_{n,k+1})$. To justify this, assume $\mu = s_{nk}$ for some $n,k$, i.e., $\phase{\mu}=\alpha_n + (k-\frac{1}{2})\omega$. With this, the lemma in (\ref{eqn:lemma2}) results in not one but two possible solutions, $\theta_{n,1}^*=k\omega$ and $\theta_{n,2}^*=(k-1)\omega$, for which $\theta_{n,1}^* + \alpha_n -\phase{\mu} = \frac{\omega}{2}$ and $\theta_{n,2}^* + \alpha_n -\phase{\mu} = -\frac{\omega}{2}$, respectively. Note that $\theta_{n,1}^*$ and $\theta_{n,2}^*$ cannot make $|g|$ in (\ref{eqn:absg}) real-valued at the same time, because $\sin(\theta_{n,1}^* + \alpha_n - \phase{\mu}) = (-1)\sin(\theta_{n,2}^* + \alpha_n - \phase{\mu})$. Therefore only one of them can be the optimum selection, which we already consider separately for scenarios when $\mu \in {\rm arc} (s_{n,k-1}:s_{nk})$ and $\mu \in {\rm arc} (s_{nk}:s_{n,k+1})$, i.e., $\mu$ is right before and right after $s_{nk}$, respectively.

We note that $g$ in (\ref{eqn:eqn3}) is defined for optimal phases $(\theta_1^*,\theta_2^*,\ldots,\theta_N^*)$. The $\mu$ in (\ref{eqn:lemma2}),
which comes from the definition in (\ref{eqn:eqn4})
which follows from (\ref{eqn:eqn3}), is the optimum one. On the other hand, in the rest of
the paper, when we refer to $\mu$, it is a value we are considering in search of the optimal $\mu$.

\subsection{UNIFORM POLAR QUANTIZATION} \label{sec:upqSec}
To address the discrete constraint on the RIS phase shifts, a straightforward approach is to project the relaxed continuous solution to the closest discrete value in the discrete phase shift set $\Phi_K$. In \cite{ZSRLCL22}, the authors named the discretization process of the continuous solutions as the closest point projection (CPP). Note that, by its definition, the CPP approach can be employed over any other algorithm that gives the relaxed solutions. Thus, the time-complexity of the discrete beamforming problem with CPP is dependent on the complexity of the algorithm that gives the preliminary continuous solution. For example, in \cite{XDMWQ22}, the authors used Discrete Manifold Optimization (Discrete-Manopt), which corresponds to quantization of the continuous phase shifts provided by Manifold Optimization, and the computational time complexity to achieve the quantized solutions is extremely high.

In this section, we will define a practical intuitive algorithm that can achieve a suboptimal solution, similar to the CPP approach in \cite{ZSRLCL22}, which we call uniform polar quantization (UPQ). For this purpose, we can redefine the received power maximization problem with relaxed continuous phase shifts as follows:
\begin{equation}
	\begin{aligned}
		& \underset{\mbox{\boldmath$\theta$}^{\text{cont}}}{\rm maximize\ } f_{\text{rx}}({\mbox{\boldmath$\theta$}^{\text{cont}}})\\
		& {\rm subject\ to\ } \theta_n^{\text{cont}} \in [0,2\pi),\ n=1, 2, \ldots, N
	\end{aligned}
	\label{eqn:eqn1rx}
\end{equation}
where
\begin{equation}
	f_{\text{rx}}({\mbox{\boldmath$\theta$}}) = \bigg|\beta_0e^{j\alpha_0}+\sum_{n=1}^N \beta_n
	e^{j(\alpha_n + \theta_n)}\bigg|^2.
	\label{eqn:frx}
\end{equation}
Note that in the magnitude of the objective term $f_{\text{rx}}({\mbox{\boldmath$\theta$}})$, we are adding complex numbers, or equivalently two-dimensional vectors on the complex-plane. Therefore, the bounds on the received power in (\ref{eqn:frx}) can simply be given as $0 \leq f_{\text{rx}}({\mbox{\boldmath$\theta$}}) \leq \left(\sum_{n=0}^{N}\beta_n\right)^2$.
In this case, assuming continuous phase shifts, the solution to the maximization problem would be to find $\theta_n^{\text{cont}}$ such that
\begin{equation}
	\alpha_n + \theta_n^{\text{cont}} = \alpha_0,\ {\rm for}\  n=1,2,\dots,N.
	\label{eqn:findThetaCont}
\end{equation}
Therefore, by letting $\theta_n^{\text{cont}} = \alpha_0 - \alpha_n$ for $n=1,2,\dots,N$, we can select the discrete phase shifts with the following rule, which we refer to as UPQ in this paper:
\begin{equation}
	\theta_n^{\text{UPQ}} = \bigg\lfloor \frac{\alpha_0 - \alpha_n}{\omega} \bigg\rceil \times \omega,\ {\rm for}\  n=1,2,\dots,N,
	\label{eqn:upq}
\end{equation}
where $\lfloor \cdot \rceil$ is the rounding function defined as
\begin{equation}\label{eqn:round}
	\lfloor x \rceil = {\rm sgn} (x)\left\lfloor |x| + 0.5 \right\rfloor.
\end{equation}
The importance of defining UPQ for the problem in (\ref{eqn:eqn1}), or equivalently for a discrete QP with the rank constraint is to present a fair comparison in the computational complexity results, where UPQ performs surprisingly well in terms of both performance and computational complexity given full CSI. We note that, in terms of full CSI, UPQ only requires $\alpha_n$ for $n=0,1,\dots,N$, and in a real scenario, look-up tables can be employed to further simplify the beamforming process.

Similar to UPQ, there are approaches used under different names in the literature. Firstly, as mentioned, \cite{ZSRLCL22} uses a similar approach for the CPP result where the authors show performance over estimated channels. Another example is, in \cite{PYTCLWZB21}, the authors derive a measurement based beamforming algorithm that is called the ``Greedy Fast Beamforming Algorithm,'' which is based on a similar quantization approach employed for the binary case.

We note that, a secondary result of equation (\ref{eqn:findThetaCont}) is to show why the optimal discrete phase shift selection problem was originally thought to be very difficult to solve. In (\ref{eqn:findThetaCont}), $\alpha_n$ and $\alpha_0$ values are continuous. When $\theta_n^{\text{cont}}$ are constrained by the set $\Phi_K$, it practically prevents a closed-form solution for $\theta_n$ to be available, because in (\ref{eqn:frx}) each $\theta_n$ is related to each other. Yet, in the following sections, our derivations with the optimal algorithm show that the optimum discrete phase shift selection problem turns out to be simpler, actually.
\begin{table}[!t]
	\begin{center}
		\caption{Approximation ratio of the UPQ Algorithm to the continuous solution.}
		\label{tbl:ErxTable}
		\begin{tabular}{lccccc}
			\hline
			&$K=2$&$K=3$&$K=4$&$K=6$&$K=8$\\
			\hline
			\hline
			$E_\infty(K)$&0.4053&0.6839&0.8106&0.9119&0.9496\\
			\hline
			$1/E_\infty(2)$ & -- &2.27 dB&3.01 dB&3.52 dB&3.70 dB\\
			\hline
		\end{tabular}
	\end{center}
\end{table}

\subsubsection{Efficiency Calculation for UPQ}
In \cite{ZSRLCL22}, the authors provide a lower bound on the performance of quantization. In this section, similar to the approach in \cite{WZ20}, the performance loss in UPQ due to quantization is quantified with respect to $K$. For this purpose, the expected value of the normalized performance, i.e., $E_N(K)$, is calculated for asymptotically large $N$. For large $N$, equation (\ref{eqn:frx}) with the UPQ solution can be rewritten as
\begin{align}
	f_{\text{rx}}({\mbox{\boldmath$\theta$}^{\text{UPQ}}}) =& \bigg|\beta_0e^{j\alpha_0}+\sum_{n=1}^N \beta_n
	e^{j(\alpha_n + \theta^{\text{UPQ}}_n)}\bigg|^2 \nonumber\\
	= & \left|e^{j\alpha_0}\right|^2 \bigg|\beta_0+\sum_{n=1}^N \beta_n
	e^{j(\alpha_n+\theta^{\text{UPQ}}_n-\alpha_0)}\bigg|^2 \nonumber\\
	= & \bigg|\beta_0+\sum_{n=1}^N \beta_n
	e^{j(\theta^{\text{UPQ}}_n-\theta^{\text{cont}}_n)}\bigg|^2\nonumber\\
	\approx & \bigg|\sum_{n=1}^N \beta_n
	e^{j(\theta^{\text{UPQ}}_n-\theta^{\text{cont}}_n)}\bigg|^2, \label{eqn:frx2aprx}
\end{align}
where, for asymptotically large $N$, $\beta_0$ from the BS-UE direct link in (\ref{eqn:frx2aprx}) is practically discarded. Therefore, the received power for large $N$ can be approximated as
\begin{align}
	f_{\text{rx}}&({\mbox{\boldmath$\theta$}^{\text{UPQ}}}) \approx \bigg|\sum_{n=1}^N\beta_n e^{j(\theta^{\text{UPQ}}_n-\theta^{\text{cont}}_n)}\bigg|^2 = \sum_{n=1}^N \beta_n^2 \nonumber \\
	+& \; 2 \sum_{k=2}^N\sum_{l=1}^{k-1} \beta_k\beta_l \cos((\theta^{\text{UPQ}}_k-\theta^{\text{cont}}_k)-(\theta^{\text{UPQ}}_l-\theta^{\text{cont}}_l)). \label{eqn:frxApxCos}
\end{align}
Assume that in (\ref{eqn:frxApxCos}) all $\beta_k, \beta_l, \theta_k$, and $\theta_l$ are independent from each other. Let $\delta_i=\theta^{\text{UPQ}}_i-\theta^{\text{cont}}_i$ for $i=1,\dots,N$, so that in (\ref{eqn:frxApxCos}), the argument of the cosine is equal to $\delta_k-\delta_l$. From equation (\ref{eqn:upq}), it follows that $\delta_i \in \left[-\frac{\pi}{K},\frac{\pi}{K}\right]$. Assume $\delta_k$ and $\delta_l$ are i.i.d. uniform random variables in $\left[-\frac{\pi}{K},\frac{\pi}{K}\right]$, i.e., $\delta_k,\delta_l\sim\mathcal{U}\left[-\frac{\pi}{K},\frac{\pi}{K}\right]$, which results in $\mathbb{E}\left[\cos\left(\delta_m-\delta_n\right)\right]= \sinc^2\left(\frac{1}{K}\right)$. Therefore, the expected value of $f_{\text{rx}}({\mbox{\boldmath$\theta$}^{\text{UPQ}}})$ can be defined as
\begin{equation} \label{eq:Efrx}
	\mathbb{E}[f_{\text{rx}}({\mbox{\boldmath$\theta$}^{\text{UPQ}}})] = N\mathbb{E}[\beta_n^2] + N(N-1)\mathbb{E}[\beta_k\beta_l]\sinc^2\bigg(\frac{1}{K}\bigg).
\end{equation}
With this, we normalized the received power by the expected value of the maximum achievable power, i.e., $\mathbb{E}[(\sum_{n=1}^N \beta_n)^2] = N\mathbb{E}[\beta_n^2] + N(N-1)\mathbb{E}[\beta_k\beta_l]$, which gives
\begin{equation}
	E_{N}(K) \triangleq \frac{N\mathbb{E}[\beta_n^2] + N(N-1)\mathbb{E}[\beta_k\beta_l]\sinc^2\left(\frac{1}{K}\right)}{N\mathbb{E}[\beta_n^2] + N(N-1)\mathbb{E}[\beta_k\beta_l]}.
	\label{eqn:ENK}
\end{equation}
Taking $N \to \infty$, the expected value of the normalized performance can be represented as a function of $K$ as
\begin{equation} \label{eq:expectedPower2}
	E_\infty(K) = \sinc^2\left(\frac{1}{K}\right),
\end{equation}
which quantifies the effect of $K$, i.e., the number of available discrete phase shift selections, as an approximation to the optimal continuous solution. For example, as $K \to \infty$, i.e., the continuous case, $E_\infty(\infty) = 1$, therefore $E_\infty(K)$ also serves as an approximation ratio to the upper bound. Two examples for different selections of $K$ are given in Table \ref{tbl:ErxTable}, where it can be observed that gains from using at least $K=4$ discrete phases over $K=2$ are significant.

A final emphasis we want to make is to show that the optimal result can be achieved as a quantization of the relaxed solution. In other words, the optimal solution $\theta_n^*$ given $\mu$ using UPQ, i.e., $\theta_{n|\mu}^{\text{UPQ}}$, can be achieved by employing the UPQ algorithm with $\theta_n^{\text{cont}} = \phase{\mu} - \alpha_n$. To see this, assume $\mu$ satisfies (\ref{eqn:eqn8}). From (\ref{eqn:suffRange}), we know that $\phase{\mu} - \alpha_n \in \Big((2k-1)\frac{\pi}{K}, (2k+1)\frac{\pi}{K}\Big)$, in which case the quantization in UPQ gives
\begin{equation}\label{eqn:upqMu}
	\theta_{n|\mu}^{\text{UPQ}} = \bigg\lfloor \frac{\phase{\mu} - \alpha_n}{\omega} \bigg\rceil \times \omega = k\omega,
\end{equation}
for $n=1,\dots,N$. Note that the quantization result in (\ref{eqn:upqMu}) is compatible with Proposition~1, which proves that the optimum result is actually a quantization solution. As the two approaches (equation (\ref{eqn:lemma2}) and quantization) are akin, UPQ can provide close-to-optimal solution given full CSI knowledge. However, as $\mu$ is unknown and one needs to search for the optimal $\mu$, we present this as an insight on the problem rather than an operational idea.

Finally, although both CPP and UPQ give the quantized continuous solution without modification, they have a basic difference. CPP is applied over any continuous solution and defined by $\theta_n^{\text{CPP}} = \arg \min_{\theta_n\in \Phi_K} |\theta_n - \theta_n^{\text{cont}}|$ \cite{ZSRLCL22}. On the other hand, UPQ is specifically defined for the problem in (\ref{eqn:eqn1}) by using the rounding operator in equation (\ref{eqn:round}) to determine $\theta_n^{\text{UPQ}}$ directly from $\alpha_n$ and $\alpha_0$. Using the $\arg \min$ operator not only gives marginal insight on the problem of interest, but it can lead to incorrect results for specific values of $\theta_n \in [0,2\pi)$ and $\theta_n^{\text{cont}} \in [0,2\pi)$.

In the next section, we will define our discrete phase shift selection algorithm that guarantees the global optimal solution. We further improve it in the sequel to converge with the least number of steps, and achieve significantly lower computational complexity.

\section{A NEW ALGORITHM}
\begin{algorithm}[!t]
\caption{Update for \cite[Algorithm~1]{b1}}\label{alg:alg2}
\begin{algorithmic}[1]
\State {\bf Initialization:} Compute 
$s_{nk}=e^{j(\alpha_n + (k -0.5)\omega)}$ for $n=1,2,\ldots,N$ and
$k=1,2,\ldots,K$.
\State Eliminate duplicates among $s_{nk}$ and sort to get $e^{j\lambda_l}$
such that $0\le \lambda_1 < \lambda_2 < \cdots < \lambda_L < 2\pi.$
\State Let, for $l = 1,2,\ldots,L,$ ${\cal N} (\lambda_l) =
\{ n | \phase{s_{nk}} = \lambda_l \}.$
\State 
Set $\phase{\mu} = 0$. For $n=1,2,\ldots,N$, calculate
$\theta_n = \arg\max_{\theta_n\in\Phi_K} \cos(\theta_n + \alpha_n - \phase{\mu})$.
\State Set $g_0 = h_0 + \sum_{n=1}^N h_ne^{j\theta_n}$, ${\tt absgmax} = |g_0|$.
\For{$l = 1, 2, \ldots, L-1$}
\State For each $n\in{\cal N}(\lambda_l)$, let $(\theta_n + \omega \leftarrow \theta_n) \;{\rm mod}\;\Phi_K$.
\State Let
\[
g_l = g_{l-1} + \sum_{n\in {\cal N}(\lambda_l)} h_n \big(e^{j\theta_n} - e^{j(\theta_n - \omega) \;{\rm mod}\; \Phi_K}\big)
\]
\If{$|g_l| > {\tt absgmax}$}
\State Let ${\tt absgmax} = |g_l|$
\State Store $\theta_n$ for $n=1,2,\ldots,N$
\EndIf
\EndFor
\State Read out $\theta_n^*$ as the stored $\theta_n$, $n=1,2,\ldots,N$.
\end{algorithmic}
\end{algorithm}

Reference \cite[Algorithm~1]{b1}, as published, employs the criterion
\begin{equation}
\theta_n^* = \arg \min_{\theta_n\in \Phi_K} |(\theta_n + \alpha_n -\phase{\mu})\; {\rm mod}\; 2\pi|
\end{equation}
instead of (\ref{eqn:lemma2})\footnote{In this paper, we define the ${\rm mod}$
function (the modulus function or the modulo operation) $x\; {\rm mod}\;y$ as the remainder after the
dividend $x>0$ is divided by the divisor $y>0$. We write it as $x\;{\rm mod}\; y$, $x\; (mod\; y)$, or
${\rm mod}\; (x, y)$. For $x<0$ and $y>0,$ we use the convention that the remainder should always
be the smallest such nonnegative number.}. We now specify Algorithm~1 as an alternative to
\cite[Algorithm 1]{b1}\footnote{In Algorithm~\ref{alg:alg2} we define $(\theta \pm \omega)\,{\rm mod}\,\Phi_K$ as follows. First note that the two
sets $\{0,\omega,2\omega,\dots,(K-1)\omega\}$ and $\{\omega,2\omega,3\omega,\ldots,K\omega\}$ have the same
members since $\omega = 2\pi/K$. Then, $(\theta + \omega)\,{\rm mod}\,\Phi_K$ can be defined as
$(\theta \pm \omega)\,{\rm mod}\,\Phi_K \triangleq ((k \pm 1)\,{\rm mod}\,K)\,\omega$.}.

\begin{figure*}[!t]
\centering
\begin{minipage}{0.48\textwidth}
\centering
\includegraphics[width=1.0\textwidth]{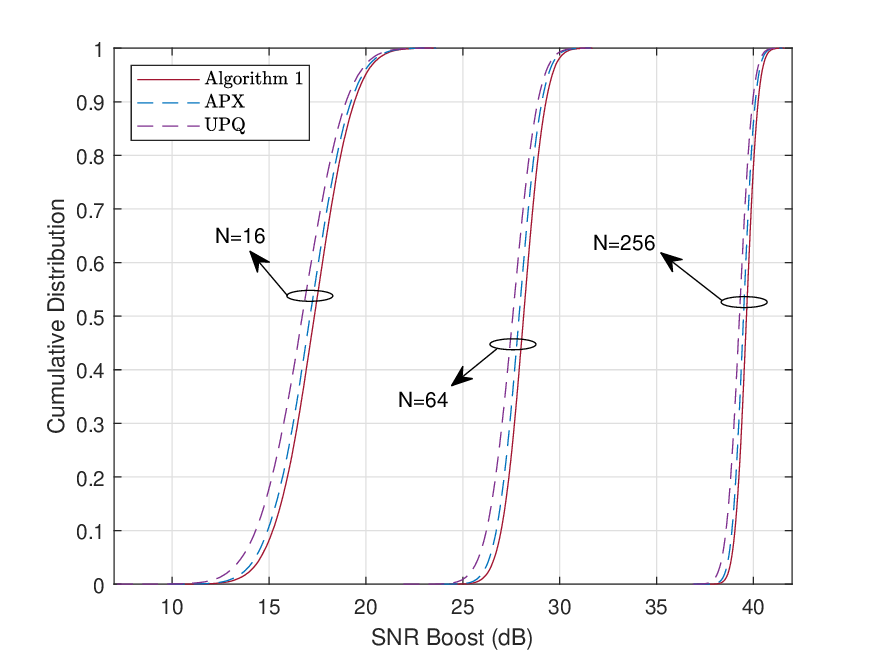}
\caption{CDF plots for SNR Boost \cite{b1} with Uniform Polar Quantization (UPQ), Algorithm~1, and Approximation (APX) Algorithm \cite{ZSRLCL22}, $K=2$.}
 \label{fig:SNRBoostK2}
\end{minipage}%
\hspace{0.03\textwidth}
\begin{minipage}{0.48\textwidth}
\centering
\includegraphics[width=1.0\textwidth]{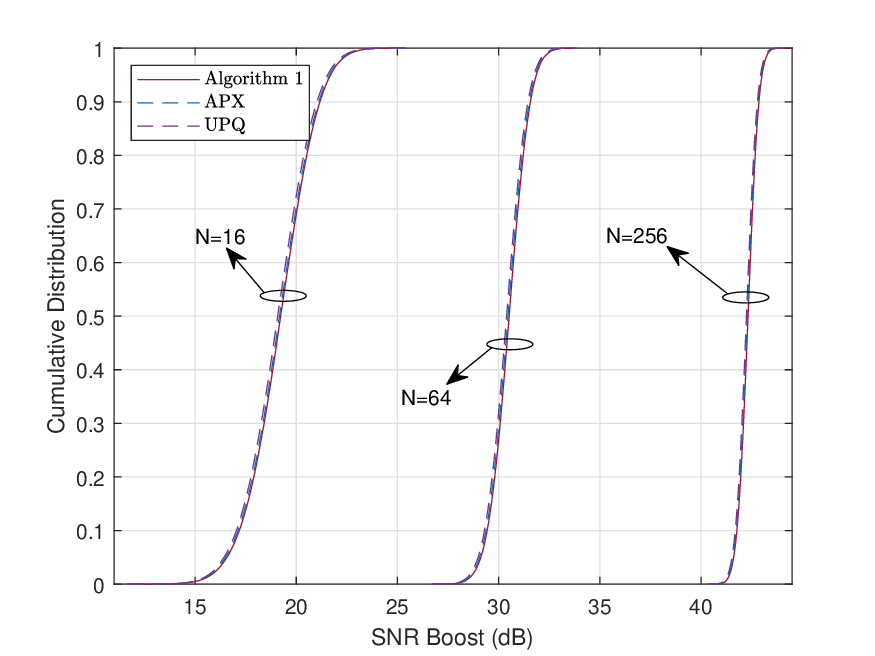}
\caption{CDF plots for SNR Boost \cite{b1} with Uniform Polar Quantization (UPQ), Algorithm~1, and Approximation (APX) Algorithm \cite{ZSRLCL22}, $K=4$.}
 \label{fig:SNRBoostK4}
\end{minipage}
\end{figure*}
\begin{figure*}[!t]
\centering
\begin{minipage}{0.48\textwidth}
\centering
\includegraphics[width=1.0\textwidth]{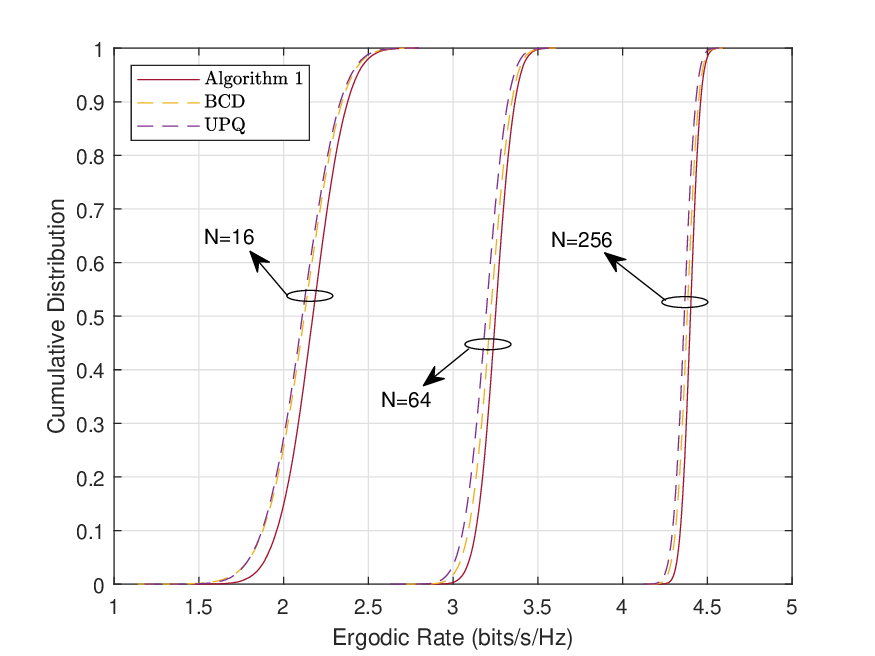}
\caption{CDF plots for Ergodic Rate (\ref{eq:ergodicRate}) with Uniform Polar Quantization (UPQ), Algorithm~1, and block coordinate descend (BCD) \cite{AZWY20}, $K=2$.}
 \label{fig:ErgodicRateK2}
\end{minipage}%
\hspace{0.03\textwidth}
\begin{minipage}{0.48\textwidth}
\centering
\includegraphics[width=1.0\textwidth]{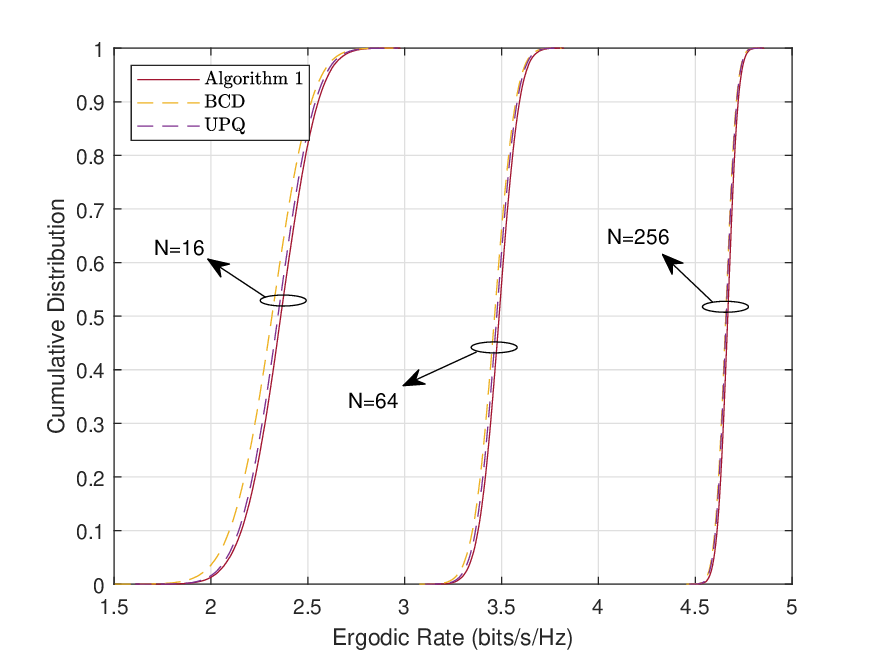}
\caption{CDF plots for Ergodic Rate (\ref{eq:ergodicRate}) with Uniform Polar Quantization (UPQ), Algorithm~1, and block coordinate descend (BCD) \cite{AZWY20}, $K=4$.}
 \label{fig:ErgodicRateK4}
\end{minipage}
\end{figure*}
\begin{figure*}[!t]
	\centering
	\begin{minipage}{0.48\textwidth}
		\centering
		\includegraphics[width=1.0\textwidth]{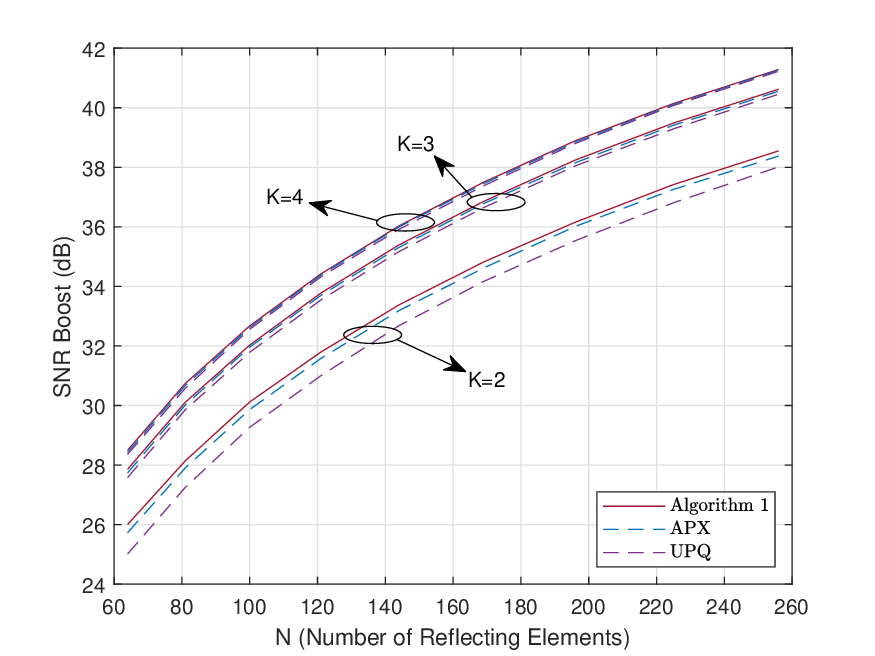}
		\caption{1st percentile SNR Boost results vs. $N$, for $K\in\{2,3,4\}$.}
		\label{fig:percentile}
	\end{minipage}%
	\hspace{0.03\textwidth}
	\begin{minipage}{0.48\textwidth}
		\centering
    \vspace{4mm}
		\includegraphics[width=1.0\textwidth]{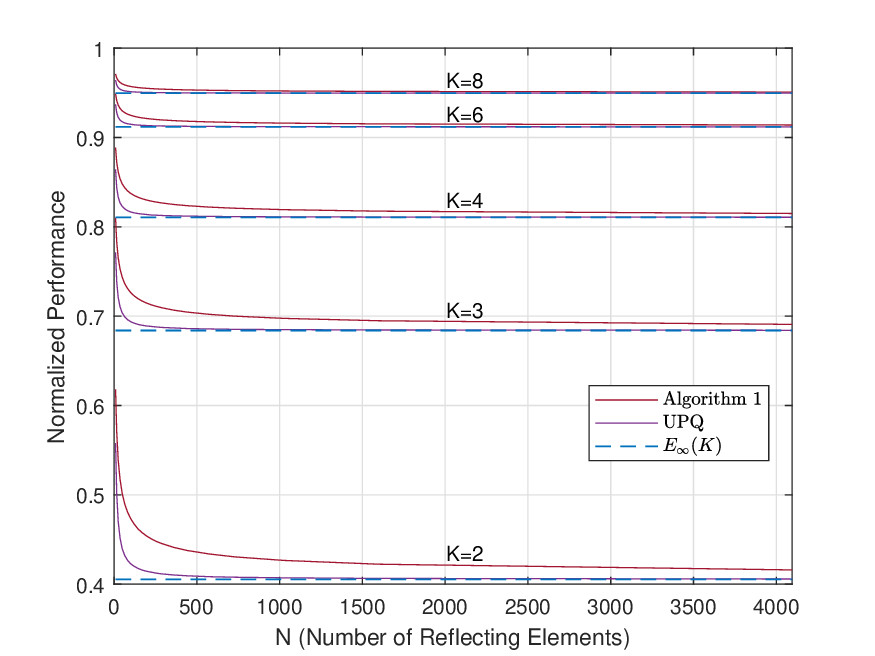}
		\caption{Normalized Performance results vs. $N$, for $\kappa=10$ and $K\in\{2,3,4,6,8\}$.}
		\label{fig:normPerf}
	\end{minipage}
\end{figure*}

We present the cumulative distribution function (CDF) results for SNR Boost\cite{b1} in Fig.~\ref{fig:SNRBoostK2} for $K=2$, and in Fig.~\ref{fig:SNRBoostK4} for $K=4$. Similarly, the ergodic rate results are presented in Fig.~\ref{fig:ErgodicRateK2} and \ref{fig:ErgodicRateK4}, respectively. The CDF results are presented for $N=16,$ $64,$ and $256,$ using the average of 10,000 realizations of the channel model defined in Section \ref{sec:channelmodel} with $\kappa=0$, where we employed UPQ, Algorithm~1, and the Approximation (APX) algorithm from \cite{ZSRLCL22}, where all algorithms ran over the same realization in each step. For the ergodic rate results, instead of APX algorithm, we employed the block coordinate descend (BCD) \cite{AZWY20} as a typical benchmark where the phase shifts are optimized for each RIS element at a time. Although the gains are not large, especially with $K=4$ in Fig.~\ref{fig:SNRBoostK4}, these two figures serve as a verification of the optimality of Algorithm~1, which we already know from the analysis presented in this paper. Besides, in Fig. \ref{fig:percentile}, we focus on the low SNR Boost regime and compare the 1st percentile results. For $K=2$, while there is only about 1 dB loss of using UPQ, it cannot be recovered by increasing the number of RIS elements. On the other hand, for $K \geq 4$, both APX and UPQ can provide close-to-optimal results, with UPQ being quite efficient in terms of complexity, which we discuss in Section~\ref{sec:compcomp}.

Finally, the normalized received power results are calculated by $|\beta_0e^{j\alpha_0}+\sum_{n=1}^N \beta_n
e^{j(\alpha_n + \theta_n)}|^2 / (\sum_{n=0}^{N}\beta_n)^2$ and plotted in Fig. \ref{fig:normPerf} for $\kappa=10$. The figure verifies that the expected value approximation for large $N$, i.e., $E_\infty(K)$ results in Table \ref{tbl:ErxTable}, fall in line with the numerical results.

In the Appendix, we discuss an alternative way to initialize Algorithm~1, which significantly reduces the computational complexity. 
We will use this technique in initializing Algorithm~2 and Algorithm~3 in the sequel.

\section{ALGORITHM CONVERGENCE: TOWARDS TWO NEW ALGORITHMS}
We will now show the periodicity in the update rule in Algorithm~1, i.e.,
in ${\cal N}(\lambda_l)$. With this observation, we will
prove that Algorithm~1 takes $N$ or fewer steps to converge,
as opposed to the statement that \cite[Algorithm~1]{b1} takes $KN$ or $2N$ steps on
average \cite{b1}. This will result in two new versions of the algorithm, i.e., Algorithm~2
and Algorithm~3, with a simple elementwise update rule, in the sequel.
Towards this end, we first make the following statement.

{\em Claim 1:\/} As in (\ref{eqn:eqn12}), set $s_{nk}=e^{j(\alpha_n+(k-\frac{1}{2})\frac{2\pi}{K})}$,
$n=1,2,\ldots, N,$ $k=1,2,\ldots,K$, $\alpha_n\in [0,2\pi ).$ Let $\lambda_l = \phase{s_{nk}}$ such
that $0\le \lambda_1 < \lambda_2 < \cdots < \lambda_L <2\pi$. Let ${\cal N}(\lambda_l)=\{n|\lambda_l = \phase{s_{nk}}\}$.
Assuming for now that $|{\cal N}(\lambda_l)| = 1$, $l=1,2\ldots,L=NK$, which we will relax in the sequel,
we claim that ${\cal N}(\lambda_{l'}) = {\cal N}(\lambda_{l'+N})$ for $l'=1,2,\ldots,N(K-1).$

To prove {\em Claim 1,\/} we will first introduce {\em Claim~2\/} and prove it.

{\em Claim 2:\/} Without loss of generality, we can assume that $\alpha_n < \frac{2\pi}{K},$ $n=1,2,\ldots,N$.

{\em Proof of Claim 2:\/} Suppose that for some $n$, we have $\frac{2\pi}{K}\cdot m \le \alpha_n < \frac{2\pi}{K}\cdot (m+1),$
$m=1, 2, \ldots, K-1$. Let $\beta_n \triangleq \alpha_n - m\cdot \frac{2\pi}{K},$ so that $\beta_n< \frac{2\pi}{K}$.
We will write below each phase value in (\ref{eqn:eqn12}), $\alpha_n + \frac{(2k-1)\pi}{K}$, for $k=1,2,\ldots,K$ (note that
the $({\rm mod}\; 2\pi)$ notation below applies to both sides of the equation).
\begin{itemize}
\item $k=1\hspace{-1mm}:$\\ $\alpha_n+\frac{\pi}{K} = \beta_n+\frac{(2m+1)\pi}{K} \ ({\rm mod}\; 2\pi)$\\
\hspace{15mm}where $\Big(\frac{(2m+1)\pi}{K}\Big)_{m=1}^{K-1} = \Big\{\frac{3\pi}{K},\frac{5\pi}{K},\ldots,\frac{(2K-1)\pi}{K}\Big\},$
\item $k=2\hspace{-1mm}:$\\ $\alpha_n+\frac{3\pi}{K} = \beta_n + \frac{(2m+3)\pi}{K}\ ({\rm mod}\; 2\pi)\\ =\left\{\begin{tabular}{ll}
$\beta_n +\frac{(2m+3)\pi}{K}\ ({\rm mod}\; 2\pi),$&$m \le K-2$\\
$\beta_n +\frac{\pi}{K}\ ({\rm mod}\; 2\pi),$&$m = K-1$
\end{tabular}\right.$\\
\hspace{15mm}where $\Big(\frac{(2m+3)\pi}{K}\Big)_{m=1}^{K-2} = \Big\{\frac{5\pi}{K},\frac{7\pi}{K},\ldots,\frac{(2K-1)\pi}{K}\Big\},$\\
$\vdots$
\item $k=K\hspace{-1mm}:$\\ $\alpha_n+\frac{(2K-1)\pi}{K} = \beta_n +\frac{(2m+2K-1)\pi}{K}\ ({\rm mod}\; 2\pi)\\
\mbox{\hspace{20.8mm}}=\beta_n+\frac{(2m-1)\pi}{K}\ ({\rm mod}\; 2\pi)$ \\
\hspace{15mm}where $\Big(\frac{(2m-1)\pi}{K}\Big)_{m=1}^{K-1} = \Big\{\frac{\pi}{K},\frac{3\pi}{K},\ldots,\frac{(2K-3)\pi}{K}\Big\}.$
\end{itemize}
Thus, if there is an $\alpha_n\ge \frac{2\pi}{K}$ to generate $K$ phase values, there is always a $\beta_n$, $\beta_n< \frac{2\pi}{K}$
with which one can generate the same $K$ phase values in a similar fashion. Therefore, in order to
prove Claim 1, one can work with the assumption that $\alpha_n<\frac{2\pi}{K},$ for $n=1,2,\ldots, N$.\hfill$\blacksquare$

{\em Proof of Claim 1:\/} 
Assuming $0\le \alpha_1 < \alpha_2 < \cdots < \alpha_N<\frac{2\pi}{K},$ 
without loss of generality,
we will now show that
${\cal N}(\lambda_{l'}) = {\cal N}(\lambda_{l'+N})$ for $l'=1,2,\ldots,N(K-1).$ For this, there are $N+1$ cases to consider.

{\em Case 0:\/} In this case, we assume $\alpha_n<\frac{\pi}{K},$ $n=1,2,\ldots,N$. We write all possible values of
$\phase{s_{nk}}$ as follows.
\begin{flushleft}
$n=1\hspace{-1mm}:$
\end{flushleft}
\vspace{-5mm}
\begin{equation}
\phase{s_{1k}}\in \textstyle\left\{ \alpha_1+ \frac{\pi}{K}, \alpha_1 +\frac{3\pi}{K}, \ldots , \alpha_1 + \frac{(2K-1)\pi}{K}\right\},
\label{eqn:eqn28}
\end{equation}
\begin{flushleft}
$n=2\hspace{-1mm}:$
\end{flushleft}
\vspace{-5mm}
\begin{equation}
\phase{s_{2k}}\in \textstyle\left\{ \alpha_2+ \frac{\pi}{K}, \alpha_2 +\frac{3\pi}{K}, \ldots , \alpha_2 + \frac{(2K-1)\pi}{K}\right\},
\label{eqn:eqn29}
\end{equation}
\hspace{7mm}$\vdots$
\begin{flushleft}
$n=N\hspace{-1mm}:$
\end{flushleft}
\vspace{-5mm}
\begin{equation}
\hspace{4mm}\phase{s_{Nk}}\in \textstyle\left\{ \alpha_N+ \frac{\pi}{K}, \alpha_N +\frac{3\pi}{K}, \ldots , \alpha_N + \frac{(2K-1)\pi}{K}\right\}.
\label{eqn:eqn30}
\end{equation}
Sorting (\ref{eqn:eqn28})--(\ref{eqn:eqn30}), we have
\begin{equation}
\begin{aligned}
\hspace{2mm} & \textstyle \alpha_1 + \frac{\pi}{K} < \alpha_2 + \frac{\pi}{K} < \cdots < \alpha_N + \frac{\pi}{K} \\
& \!\!\!\!\!\! < \textstyle \alpha_1 + \frac{3\pi}{K} < \alpha_2 + \frac{3\pi}{K} < \cdots < \alpha_N + \frac{3\pi}{K} \\
& \!\!\!\!\!\!\hspace{2mm}\vdots\\
& \!\!\!\!\!\! < \textstyle \alpha_1 + \frac{(2K-1)\pi}{K} < \alpha_2 + \frac{(2K-1)\pi}{K} < \cdots < \alpha_N + \frac{(2K-1)\pi}{K} .\nonumber
\end{aligned}
\end{equation}
Thus,
\begin{align*}
\big({\cal N}(\lambda_l)\big)_{l=1}^{L=NK} = \Big\{ \underbrace{1, 2, \ldots, N}_{1}, & \underbrace{1, 2, \ldots, N}_{2},\nonumber\\
& \hspace{5mm}\ldots,\underbrace{1, 2, \ldots, N}_{K}\Big\}.
\end{align*}
Therefore, for Case 0 and for $l'=1,2,\ldots,N(K-1),$ we have ${\cal N}(\lambda_{l'}) = {\cal N}(\lambda_{l'+N})$.

There are $N$ remaining cases. We will discuss these cases as {\em Case $i$\/} where $i = 1, 2 , \ldots , N$.

{\em Case $i, (i=1,2,\ldots, N)\!\! :$\/} In {\em Case $i$,\/} we have $i$ occurrences of $\alpha_n > \frac{\pi}{K}$ as
follows.
\begin{equation}
\begin{aligned}
0 & \le \alpha_1 < \alpha_2 < \cdots < \alpha_{N-i} < \frac{\pi}{K} \\
& \le \alpha_{N-i+1} < \cdots < \alpha_N < \frac{2\pi}{K}.
\label{eqn:eqn34}
\end{aligned}
\end{equation}
We write all possible values of $\phase{s_{nk}}$ as follows.
\begin{flushleft}
$n=1\hspace{-1mm}:$
\end{flushleft}
\vspace{-5mm}
\begin{equation}
\begin{aligned}
\hspace{-2mm}\phase{s_{1k}}\in \textstyle\Big\{ \alpha_1+ & \frac{\pi}{K}, \alpha_1 +\frac{3\pi}{K}, \ldots , \\
& \alpha_1 + \frac{(2K-3)\pi}{K}, \alpha_1 + \frac{(2K-1)\pi}{K}\Big\},
\end{aligned}
\label{eqn:eqn35}
\end{equation}
$\hspace{5mm}\vdots$
\begin{flushleft}
$n=N-i\hspace{-1mm}:$
\end{flushleft}
\vspace{-5mm}
\begin{equation}
\begin{aligned}
\phase{s_{(N-i)k}}\in \textstyle\Big\{ & \alpha_{N-i}+ \frac{\pi}{K}, \alpha_{N-i} +\frac{3\pi}{K}, \ldots , \\
 & \hspace{-10mm}\alpha_{N-i} + \frac{(2K-3)\pi}{K}, \alpha_{N-i} + \frac{(2K-1)\pi}{K}\Big\},
\end{aligned}
\label{eqn:eqn36}
\end{equation}
\begin{flushleft}
$n=N-i+1\hspace{-1mm}:$
\end{flushleft}
\vspace{-5mm}
\begin{equation}
\begin{aligned}
\phase{s_{(N-i+1)k}}\in \textstyle\Big\{ & \alpha_{N-i+1}+ \frac{\pi}{K}, \alpha_{N-i+1} +\frac{3\pi}{K}, \ldots , \\
 & \hspace{-10mm}\alpha_{N-i+1} + \frac{(2K-3)\pi}{K}, \alpha_{N-i+1} - \frac{\pi}{K}\Big\},
\end{aligned}
\label{eqn:eqn37}
\end{equation}
\begin{flushleft}
$n=N-i+2\hspace{-1mm}:$
\end{flushleft}
\vspace{-5mm}
\begin{equation}
\begin{aligned}
\phase{s_{(N-i+2)k}}\in \textstyle\Big\{ & \alpha_{N-i+2}+ \frac{\pi}{K}, \alpha_{N-i+2} +\frac{3\pi}{K}, \ldots , \\
 & \hspace{-10mm}\alpha_{N-i+2} + \frac{(2K-3)\pi}{K}, \alpha_{N-i+2} - \frac{\pi}{K}\Big\},
\end{aligned}
\label{eqn:eqn38}
\end{equation}
$\hspace{5mm}\vdots$
\begin{flushleft}
$n=N\hspace{-1mm}:$
\end{flushleft}
\vspace{-5mm}
\begin{equation}
\begin{aligned}
\hspace{-2mm}\phase{s_{Nk}}\in \Big\{ \alpha_N+ \frac{\pi}{K}, & \alpha_N +\frac{3\pi}{K}, \ldots , \\
 & \alpha_N + \frac{(2K-3)\pi}{K}, \alpha_N - \frac{\pi}{K}\Big\}.
\end{aligned}
\label{eqn:eqn39}
\end{equation}
Sorting (\ref{eqn:eqn35})--(\ref{eqn:eqn39}), we have
\begin{equation*}
\alpha_{N-i+1} - \frac{\pi}{K} < \cdots < \alpha_N - \frac{\pi}{K} \hspace{27mm}
\end{equation*}
\begin{equation*}
\hspace{30mm}< \alpha_1 +\frac{\pi}{K} < \cdots < \alpha_{N-i} + \frac{\pi}{K}
\end{equation*}
\begin{equation*}
< \alpha_{N-i+1} + \frac{\pi}{K} < \cdots < \alpha_N + \frac{\pi}{K} \hspace{30mm}
\end{equation*}
\begin{equation*}
\hspace{30mm}< \alpha_1 +\frac{3\pi}{K} < \cdots < \alpha_{N-i} + \frac{3\pi}{K}
\end{equation*}
\begin{equation*}
< \alpha_{N-i+1} + \frac{3\pi}{K} < \cdots < \alpha_N + \frac{3\pi}{K} \hspace{30mm}
\end{equation*}
\begin{equation*}
\hspace{30mm}< \alpha_1 +\frac{5\pi}{K} < \cdots < \alpha_{N-i} + \frac{5\pi}{K}
\end{equation*}
\begin{equation*}
\hspace{-50mm}\vdots
\end{equation*}
\begin{equation*}
< \alpha_{N-i+1} + \frac{(2K-3)\pi}{K} < \cdots < \alpha_N + \frac{(2K-3)\pi}{K} \hspace{1mm}
\end{equation*}
\begin{equation*}
\hspace{13mm}< \alpha_1 +\frac{(2K-1)\pi}{K} < \cdots < \alpha_{N-i} + \frac{(2K-1)\pi}{K}.
\end{equation*}
Thus,
\begin{align}
\big({\cal N}(\lambda_l)\big) & _{l=1}^{L=NK} = \nonumber\\
\Big\{ & \underbrace{N-i+1,\ldots,N,1, 2, \ldots, N-i}_{1},\nonumber\\
& \underbrace{N-i+1,\ldots,N,1, 2, \ldots, N-i}_{2}, \label{eqn:eqn387}\\
& \hspace{24.5mm}\vdots \nonumber\\
& \underbrace{N-i+1,\ldots,N,1, 2, \ldots, N-i}_{K}\Big\}\nonumber
\end{align}
for $i=1, 2 , \ldots, N$.
Therefore, for Case $i$, $i=1,2,\ldots,N$, and for $l'=1,2,\ldots,N(K-1),$ we have ${\cal N}(\lambda_{l'}) = {\cal N}(\lambda_{l'+N})$.
With this, Claim 1 is proved.
\hfill$\blacksquare$

\section{$N$ STEPS SUFFICE WHEN $|{\cal N}(\lambda_l)|=1$ FOR ALL $l$}\label{sec:nsteps}
\begin{algorithm}[!t]
\caption{Simplified Algorithm 1 with $|{\cal N}(\lambda_l)| =1$ for all $l$}\label{alg:alg31}
\begin{algorithmic}[1]
\State {\bf Initialization:} Set $\phase{\mu}=\alpha_0 - \frac{\pi}{K}$ 
\State Compute $\varphi_n = (\alpha_n - \alpha_0)$ $\mathrm{mod}\, \frac{2\pi}{K}$, $n=1,2,\ldots,N$ 
\State Sort $\varphi_n$ such that $0 \le \varphi_{n_1} < \varphi_{n_2} < \cdots < \varphi_{n_N} <\frac{2\pi}{K}$
\State Set $\theta_n = {\rm arg max}_{\theta_{n}'\in \Phi_K} \cos(\theta_{n}' +\alpha_n - \phase{\mu})$, store $\theta_n$, $n=1,2,\ldots,N$
\State Set $g_0 = h_0 + \sum_{n=1}^N h_ne^{j\theta_n}$, ${\tt absgmax} = |g_0|$
\For{$l = 1, 2, \ldots, N$}
\State Let $(\theta_{n_l} + \omega \leftarrow \theta_{n_l}) \;\mathrm{mod}\, \Phi_K$
\State Let
\[
g_l = g_{l-1} + h_{n_l} \left(e^{j\theta_{n_l}} - e^{j(\theta_{n_l} - \omega)\, \mathrm{mod}\, \Phi_K}\right)
\]
\If{$|g_l| > {\tt absgmax}$}
\State Let ${\tt absgmax} = |g_l|$
\State Store updated $\theta_{n_l}$ 
\EndIf
\EndFor
\State Read out $\theta_n^*$ as the stored $\theta_n$, $n=1,2,\ldots,N$.
\end{algorithmic}
\end{algorithm}
Given $|{\cal N}(\lambda_l)|=1$ and
\begin{equation}
{\cal N}(\lambda_{l'}) = {\cal N}(\lambda_{l'+N}),\quad l'=1,2,\dots,N(K-1),
\label{eqn:eqn371}
\end{equation}
we want to show that $N$ steps will suffice for convergence.
Now, consider the main problem of maximizing $\left|h_0 + \sum_{n=1}^N h_n e^{j\theta_n}\right|$, where it is clear that our discrete
phase shift selections can only tune the second term in the absolute value. Let
\begin{equation}
g_c \triangleq \sum_{n=1}^N h_ne^{j\theta_n} = \sum_{n=1}^N \beta_n e^{j(\alpha_n+\theta_n)} .
\label{eqn:eqn395}
\end{equation}
In each step of the Algorithm~1, we define
\begin{equation}
g_{c,l} \triangleq g_l - h_0,\quad l=1,2,\ldots,L.
\label{eqn:eqn391}
\end{equation}
Note that, in (\ref{eqn:eqn391}), $h_0$, $g_l$, and $g_{c,l}$ are complex numbers, with $l$ being a generation index. From Proposition~1,
we know that
whenever $\mu$ is anywhere in ${\rm arc}(s_{nk}:s_{n,k+1})$, $\theta_n$ does not change. The angle $\theta_n$
only changes when $\mu$ changes from one arc to another, i.e.,
\begin{equation}
\mu\in {\rm arc}{(e^{j\lambda_l}:e^{j\lambda_{l+1}})} \rightarrow \mu\in {\rm arc}(e^{j\lambda_{l+1}}:e^{j\lambda_{l+2}})
\label{eqn:eqn40}
\end{equation}
in which case $\theta_n$ must be updated as
\begin{equation}
\theta_n \rightarrow \theta_n + \omega, \quad n\in {\cal N}(\lambda_{l+1}).
\label{eqn:eqn41}
\end{equation}
With (\ref{eqn:eqn40})--(\ref{eqn:eqn41}), the naive approach in Algorithm~1 gathers all possibilities for 
$g_l$ in $NK$
steps by considering all possible arcs that $\mu$ can be in.
To show that $N$ steps will suffice, we want to point out the redundancy in those $NK$ steps.
Consider any consecutive $N$ steps in Algorithm~1.
In those steps, the phase shifts will be updated as $\theta_n \rightarrow \theta_n+\omega$
with $n\in ({\cal N}(\lambda_l))_{l=l'}^{l'+N-1}$, $l'=1,2,\ldots ,N(K-1)+1$. Since we have (\ref{eqn:eqn371}),
the following must hold
\begin{equation}
({\cal N}(\lambda_l))_{l=l'}^{l'+N-1} = \{1, 2, \ldots, N\} ,
\end{equation}
which says that after {\em any\/} $N$ consecutive steps in Algorithm~1, $\theta_n \rightarrow \theta_n + \omega,$
$n = 1, 2, \ldots, N$. To proceed further, we need an intermediate result, which we discuss below.

{\em Remark:\/} Let $g_c=\sum_{n=1}^N \beta_n e^{j(\theta_n + \alpha_n)}$ be the cascaded channel term in (\ref{eqn:eqn395}).
For any angle $\theta'$, $g_c(\theta_n+\theta') = g_c(\theta_n) e^{j\theta'}$, resulting in $|g_c(\theta_1,\theta_2,\ldots,\theta_n)| = |g_c(\theta_1+\theta',\theta_2+\theta',\ldots,\theta_N+\theta')|$. This remark illustrates that the rotation by a phase shift of an arbitrary angle $\theta'$ does not change the received power. Therefore, if $\theta_n \rightarrow \theta_n + \omega$ with $n=1,2,\ldots,N$,
\begin{equation}
|g_{c,l'}| = |g_{c,l'+N}|,\quad l'=1,2,\ldots, N(K-1)
\label{eqn:eqn43}
\end{equation}
must be true. Therefore, among the $NK$ possibilities in Algorithm~1, there are only $N$ unique values of $|g_c|$.
Consequently, as the algorithm is tuning $g_c$ to maximize $|g_c+h_0|$, it is sufficient to consider $N$ arcs that are closest
to $h_0$.

The algorithm to implement when $| {\cal N}(\lambda_l) | = 1$ for $l=1,2,\ldots,N$  is given under Algorithm~\ref{alg:alg31}.
The initialization technique introduced in the Appendix for Algorithm~1 is employed in Algorithm~2.
\section{FEWER THAN $N$ STEPS SUFFICE WHEN $|{\cal N}(\lambda_l)| > 1$ FOR SOME $l$}
With ${\cal N}(\lambda_l)$ given in Claim~1, for $|{\cal N}(\lambda_l)|
>1$ to be true for some $l$, consider a repetition among $s_{nk}$, i.e., assume there are $n_1,$ $n_2,$ $k_1,$ and $k_2$ such that
$s_{n_1,k_1} = s_{n_2,k_2}$, i.e.,
\begin{equation}
\begin{aligned}
& \left( \alpha_{n_1}+ \frac{(2k_1-1)\pi}{K}\right) \;{\rm mod}\; 2\pi = \\
& \quad \quad \quad \quad \quad \left( \alpha_{n_2}+\frac{(2k_2-1)\pi}{K}\right) \;{\rm mod}\; 2\pi .
\end{aligned}
\label{eqn:eqn44}
\end{equation}
Equation (\ref{eqn:eqn44}) is possible only if $\beta_{n_1}=\beta_{n_2}$ as $\beta_n$ are defined in Claim~1. Therefore,
all $K$ phase values represented by $\beta_{n_1}$ and $\beta_{n_2}$ must be equal, meaning there is an $N'$ such that
$M=(N-N')K$. Consequently, the problem of sorting $s_{nk}$ according to their phase values with $0\le \alpha_1 <
\alpha_2 < \cdots < \alpha_N < \frac{2\pi}{K}$ reduces to $0 \le \gamma_1 < \gamma_2 < \cdots <
\gamma_{\frac{M}{K}=N-N'} < \frac{2\pi}{K}$ for
\begin{equation}
s_{mk} = e^{j(\gamma_m+(k-\frac{1}{2})\frac{2\pi}{K})}, m=1, 2, \ldots ,\frac{M}{K}, k = 1, 2, \ldots, K
\label{eqn:eqn47}
\end{equation}
where $\gamma_m = \min \{\alpha_{n_m,1},\alpha_{n_m,2},\ldots,\alpha_{n_m,G_m}\}$ such that $\beta_{n_m,1}=\beta_{n_m,2}
= \cdots = \beta_{n_m,G_m}$. 
So, this time, there are $\frac{M}{K}+1 = N - N' + 1$ many cases.

For unique $s_{mk}$, let ${\cal M}(\lambda_l) = \{m|\lambda_l=\phase{s_{mk}}\}$. We know from (\ref{eqn:eqn387}) that
the following must hold
\begin{align}
\big({\cal M}(\lambda_l)\big) & _{l=1}^{M=(N-N')K} = \nonumber\\
\Big\{ & \underbrace{\frac{M}{K}-i+1,\ldots,\frac{M}{K},1, 2, \ldots, \frac{M}{K}-i}_{1},\nonumber\\
& \underbrace{\frac{M}{K}-i+1,\ldots,\frac{M}{K},1, 2, \ldots, \frac{M}{K}-i}_{2}, \\
& \hspace{26.5mm}\vdots \nonumber\\
& \underbrace{\frac{M}{K}-i+1,\ldots,\frac{M}{K},1, 2, \ldots, \frac{M}{K}-i}_{K}\Big\}\nonumber
\end{align}
for $i=1, 2 , \ldots, \frac{M}{K}$ where in each one of the $K$ groups there are $\frac{M}{K}=N-N'$ elements. To calculate
${\cal N}(\lambda_l)$, we define the following sets
\begin{equation*}
{\cal R}_m = \{ n_{m,1},n_{m,2},\ldots,n_{m,G_m} | \hspace{50mm}
\end{equation*}
\begin{equation}
\gamma_m = \min\{\alpha_{n_m,1},\alpha_{n_m,2},\ldots,\alpha_{n_m,G_m}\},\hspace{10mm}
\label{eqn:Rm}
\end{equation}
\begin{equation*}
\hspace{20mm}\beta_{n_m,1}=\beta_{n_m,2}=\cdots=\beta_{n_m,G_m}\}
\end{equation*}
where $G_m = |{\cal R}_m|$ and $\bigcup_{m=1}^\frac{M}{K}{\cal R}_m = \{1,2,\ldots,N\}$ must hold.
As a consequence, one can calculate
${\cal N}(\lambda_l) = {\cal R}_{{\cal M}(\lambda_l)}$. Therefore, the ``update loop'' in Algorithm~1
can be written as
\begin{align}
\big({\cal N}(\lambda_l)\big) & _{l=1}^{M=(N-N')K} = \nonumber\\
\Big\{ & \underbrace{{\cal R}_{\frac{M}{K}-i+1},\ldots,{\cal R}_\frac{M}{K},{\cal R}_1, {\cal R}_2, \ldots, {\cal R}_{\frac{M}{K}-i}}_{1},\nonumber\\
& \underbrace{{\cal R}_{\frac{M}{K}-i+1},\ldots,{\cal R}_\frac{M}{K},{\cal R}_1, {\cal R}_2, \ldots, {\cal R}_{\frac{M}{K}-i}}_{2},\label{eqn:eqn46}\\
& \hspace{26.5mm}\vdots \nonumber\\
& \underbrace{{\cal R}_{\frac{M}{K}-i+1},\ldots,{\cal R}_\frac{M}{K},{\cal R}_1, {\cal R}_2, \ldots, {\cal R}_{\frac{M}{K}-i}}_{K}\Big\}\nonumber
\end{align}
where the periodicity in the update rule still holds in (\ref{eqn:eqn46}), i.e., ${\cal N}(\lambda_{l'})={\cal N}(\lambda_{l'+\frac{M}{K}}),$ $l'=
1,2,\ldots,\frac{M}{K}(K-1)$. With the new update rule, after any $\frac{M}{K}$ consecutive steps in Algorithm~1, the phase shift selections will
be updated such that $\theta_n \rightarrow \theta_n + \omega,$ 
$n=1,2,\ldots,N$.
This will result in
\begin{equation}
|g_{c,l'}| = \left|g_{c,l'+\frac{M}{K}}\right|, \quad l'=1,2,\ldots,\frac{M}{K}(K-1).
\label{eqn:eqn50}
\end{equation}
Therefore, the sufficiency of $\frac{M}{K}=N - N'$ steps follows from (\ref{eqn:eqn43}) and the text that follows it
in Section~\ref{sec:nsteps}.

\begin{algorithm}[!t]
\caption{Simplified Algorithm~1 where $|{\cal N}(\lambda_l)| > 1$ for some $l$}\label{alg:alg32}
\begin{algorithmic}[1]
\State {\bf Initialization:} Set $\phase{\mu}=\alpha_0 - \frac{\pi}{K}$ 
\State Find $\gamma_m$ and ${\cal R}_m$ as in (\ref{eqn:eqn47}) and (\ref{eqn:Rm}), $m=1,2,\ldots,\frac{M}{K}$
\State Compute $\varphi_m = \gamma_m - \alpha_0$ $(\mathrm{mod} \frac{2\pi}{K})$, $m=1,2,\ldots,\frac{M}{K}$ 
\State Sort $\varphi_m$ such that $0 \le \varphi_{m_1} < \varphi_{m_2} < \cdots < \varphi_{m_\frac{M}{K}} <\frac{2\pi}{K}$
\State Set $\theta_n = {\rm arg max}_{\theta_{n}'\in \Phi_K} \cos(\theta_{n}' +\alpha_n - \phase{\mu})$, store $\theta_n$, $n=1,2,\ldots,N$
\State Set $g_0 = h_0 + \sum_{n=1}^N h_ne^{j\theta_n}$, ${\tt absgmax} = |g_0|$
\For{$l = 1, 2, \ldots, \frac{M}{K}=N-N'$}
\State For each $n\in{\cal R}_{m_l},$ let $(\theta_{n} + \omega \leftarrow \theta_{n}) \;\mathrm{mod}\, \Phi_K$ 
\State Let
\[
g_l = g_{l-1} + \sum_{n\in{\cal R}_{m_l}}h_{n}\left(e^{j\theta_{n}} - e^{j(\theta_{n} - \omega)\, \mathrm{mod}\, \Phi_K}\right)
\]
\If{$|g_l| > {\tt absgmax}$}
\State Let ${\tt absgmax} = |g_l|$
\State Store updated $\theta_{n}$ for $n\in {\cal R}_{m_l}$
\EndIf
\EndFor
\State Read out $\theta_n^*$ as the stored $\theta_n$, $n=1,2,\ldots,N$.
\end{algorithmic}
\end{algorithm}

Algorithm~\ref{alg:alg32} implements the technique described in this section. The initialization technique introduced in the Appendix
for Algorithm~1 is employed in Algorithm~\ref{alg:alg32}.

Note that if the BS-UE link is completely blocked, i.e., $h_0=0$, the for loop in Step~7 can end at $l=\frac{M}{K}-1=N-N'-1$, which is one fewer step to run
Algorithm~3. This is because, we can guarantee in (\ref{eqn:eqn50}) that $|g_{c,l'}| = \left|g_{c,l'+\frac{M}{K}}\right|,$ whereas we cannot say right away that
$|g_{c,l'}+h_0| = \left|g_{c,l'+\frac{M}{K}}+h_0\right|$ will be satisfied. In \cite{b1}, the authors reduce the number of steps from $KN$ to $2N$
exploiting $h_0$. In this work, with the periodicity proof, we reduce to $N$ or \textit{fewer} steps whether the direct link is blocked or not.

Step~7 in Algorithm~2 and Step~8 in Algorithm~3 are such that the phase shifts updates are restricted to just one or just a few. These steps ensure
that the running times of these algorithms are much less than those published in the literature, e.g., \cite{SBRFZT21,XDMWQ22}.

\begin{figure}[!t]
	\centering
	\includegraphics[width=0.48\textwidth]{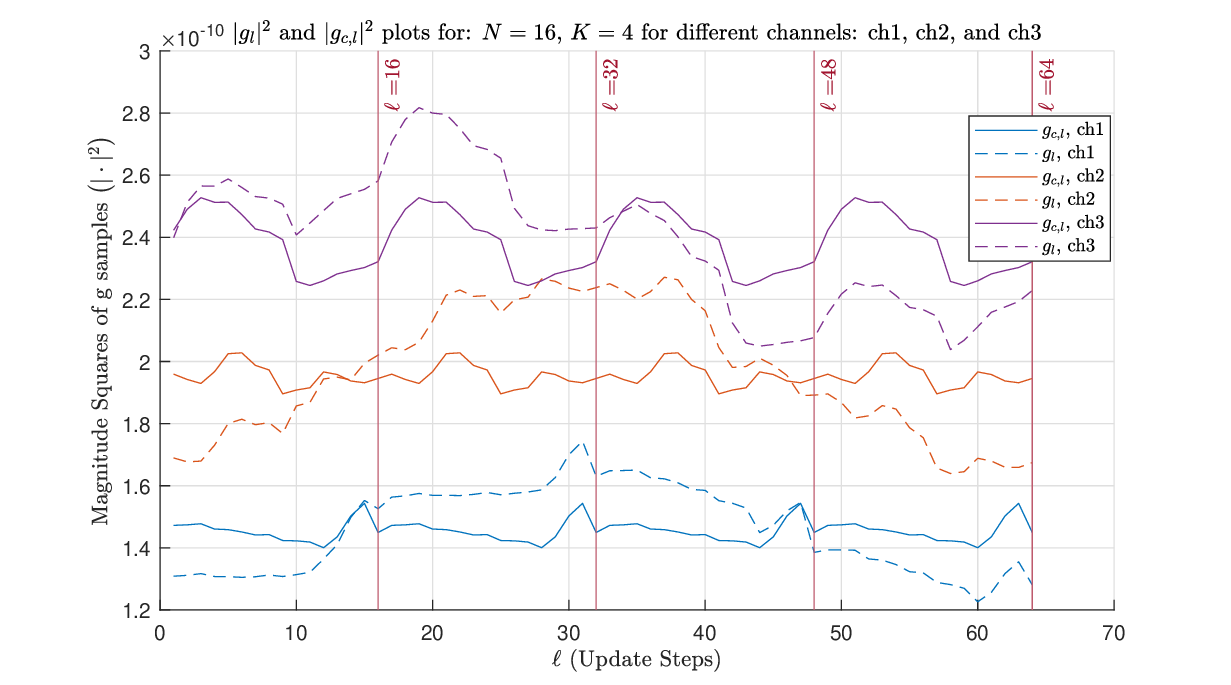}
	\caption{Variation of $|g_l|^2$ and $|g_{c,l}|^2$ with $l$.}
	\label{fig:periodicity}
\end{figure}
Figure~\ref{fig:periodicity} shows the behaviors of $|g_l|^2$ and $|g_{c,l}|^2$ against $l$ for a number of channel realizations at $N=16$ and
$K=4$. The periodicity of $|g_{c,l}|^2$ is clearly observable in this figure. It is this periodicity that we take advantage of in reducing the
number of steps for the algorithm to converge to $N$ with a simple elementwise update rule in this paper.

\section{EFFECT OF RICIAN FACTOR}
\begin{figure}[!t]
	\centering
	\includegraphics[width=0.45\textwidth]{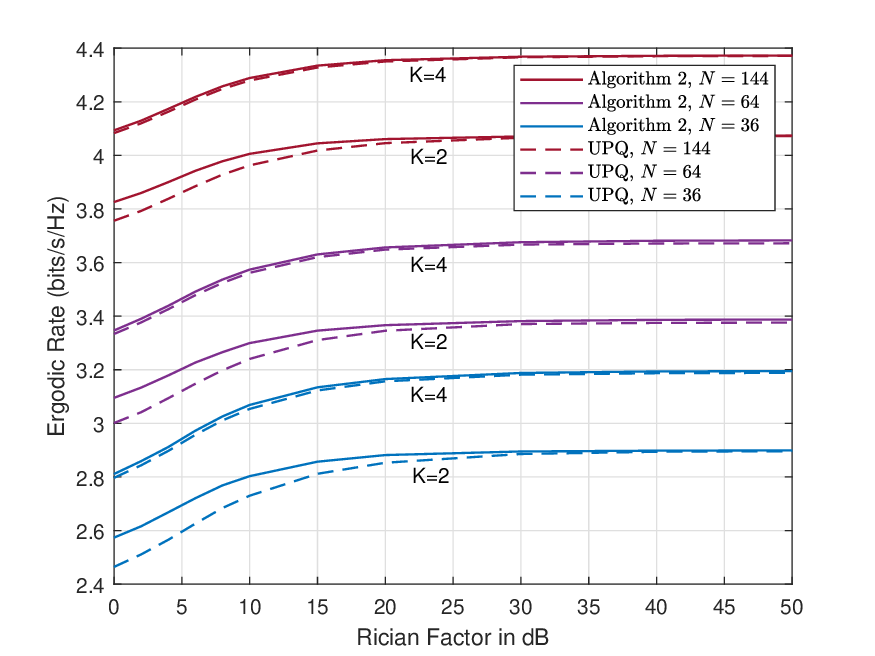}
	\caption{1st Percentile Ergodic Rate (\ref{eq:ergodicRate}) results vs. $\kappa$, for $K \in \{2,4\}$}
	\label{fig:percentileKappa}
\end{figure}

Using the given definition of the Rician channel model in equation (\ref{eq:RicianFading}), we extend the 1st percentile SNR Boost results in Fig. \ref{fig:percentile} by showing the 1st percentile ergodic rate performance results for $N \in \{36, 64, 144\}$, with different values of $\kappa$, in Fig. \ref{fig:percentileKappa}. The performance of UPQ against Algorithm~2 shows how closely UPQ can approximate the global optimum with increasing LOS gain, i.e., $\kappa$. Even with $K=2$ and lower values of $N$, the UPQ performance gets significantly closer to Algorithm~2, unveiling the potential of the simple quantization approach, i.e., UPQ. We remark that, as $\kappa$ increases, variation among each $\beta_n$ realization decreases. This results in more reliable performance provided by UPQ.

\section{EXTENSION TO MULTIUSER SCENARIO}
Similar to \cite{ZSRLCL22}, we extend our Algorithm~2 to optimize a multicast network, assuming perfect CSI with Rayleigh fading. Consider a max-min SNR problem with $U\geq2$ receivers with a transmit power of $P=30$ dBm, i.e.,
\begin{equation}
\max_{\theta_n \in \Phi_K} \min_{u} \left\{\frac{P |\beta_{0,u}e^{j\alpha_{0,u}}+\sum_{n=1}^N \beta_{n,u}
e^{j(\alpha_{n,u} + \theta_n)}|^2}{\sigma_u^2}\right\},
\label{eqn:multicastOpt}
\end{equation}
where $\sigma^2_u=-90$ dBm is the noise variance at each receive antenna, $h_{0,u} = \beta_{0,u}e^{j\alpha_{0,u}}$ is the direct channel, and $h_{n,u} = \beta_{n,u}e^{j\alpha_{n,u}}$ is the reflected channel through the $n$-th RIS element for the $u$-th receiver.

The way we extend our algorithm is as follows. While performing Algorithm~2 for user $u$, we decide the best possible solution in the for-loop of Algorithm~2 by maximizing the minimum channel gain among all users. Therefore, the for-loop of Algorithm~2 for a multicast network takes $\sum_{l=1}^N \mathcal{O}(U) = \mathcal{O}(NU)$ steps. Then, this process is repeated for each user, to select the best option among $U$ possibilities, which results in $\mathcal{O}(NU^2)$ complexity in total. We remark that the complexity of the APX algorithm in this multicast scenario is $\mathcal{O}(NU)$.

\begin{figure}[!t]
	\centering
	\includegraphics[width=0.45\textwidth]{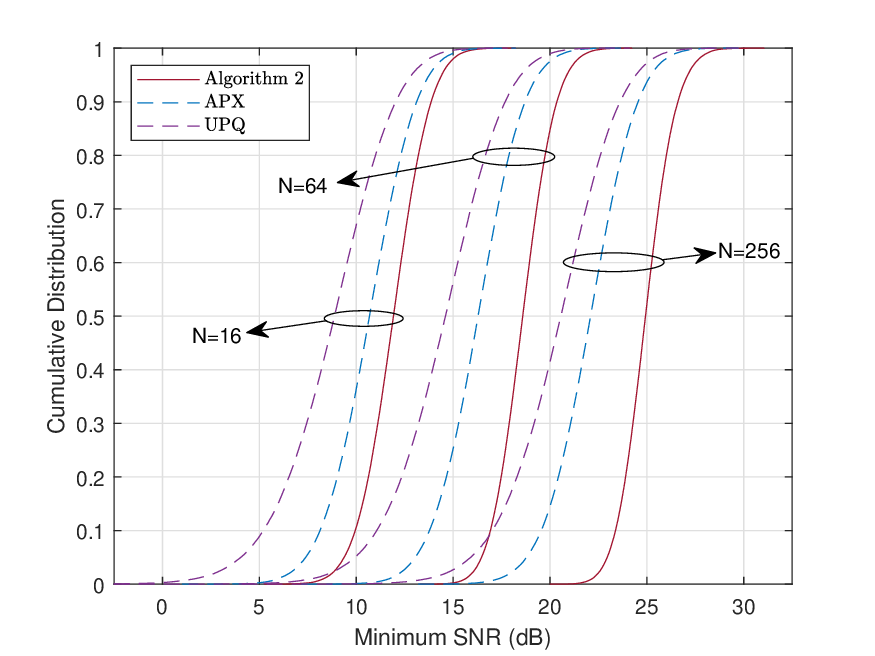}
	\caption{CDF of the minimum SNR across $U=4$ users for $K=2$ and $N \in \{16, 64, 256\}$.}
	\label{fig:multicast}
\end{figure}

The CDF plots for the minimum SNR performance of the multicast extension are given in Fig. \ref{fig:multicast} for $K=2$ and $N \in \{16, 64, 256\}$. It can be seen that Algorithm~2 can provide superior performance compared to both APX and UPQ. The average gain against UPQ is 3.0 dB and against APX is 1.3 dB for $N=16$. When $N=64$, the average gain against UPQ is 4.0 dB and against APX is 2.3 dB. Note that these gains with Algorithm~2 get larger as $N$ increases.
\section{PERFECT CSI ASSUMPTION}
In this paper, we consider a wireless communications scenario assisted by an RIS to examine the performance of our algorithms. The proposed algorithms are highly efficient compared to the current state of the literature, as shown by numerical results. One of the main contributions is the improved performance that our algorithms can achieve. With this, the proposed algorithms in this paper take $\alpha_n, n=1, \dots, N$, i.e., the channel phases, together with $\beta_n, n=1, \dots, N$ as input to give the optimal discrete beamforming solution. We assume perfect CSI while deciding the RIS phase shifts, similar to \cite{b1}. One concern can be that, since the RIS elements are passive and there is a lack of signal processing capabilities, perfect CSI for the cascaded BS-RIS-UE link practically may not be available. However, the perfect CSI assumption can relate to some scenarios with certain assumptions. For example, with a two-stage approach in RIS-aided localization systems, the passive beamforming at the RIS is performed with the available channel information from the last localization step \cite{LDDHCHB23,HWSSJ20}. Also, in experimental setups with RISs, the locations of the transmitter and receiver are known and passive beamforming is performed assuming that the location information of the UE is available \cite{PYTCLWZB21,SBRFZT21,9961233,XDMWQ22}.

As described in Section \ref{sec:channelmodel}, the channels undergo Rician fading. While the performance and numerical results are provided with $\kappa=0$ to compare with similar algorithms from the recent literature, for scenarios with the dominant line-of-sight assumption in both BS-RIS and RIS-UE, i.e., large $\kappa$, the user's direction-of-arrival (DoA) information would be sufficient to find the channel phases, similar to the geometrical optimal model proposed in \cite{MZXWZQ23}. Whereas, if the channels are strictly NLOS, i.e., $\kappa=0$ corresponding to Rayleigh fading, an example in the literature for estimating the channels in BS-RIS and RIS-UE links is to estimate the channels \cite{JD20}, which is the technique used in \cite{ZSRLCL22} for performance analysis. Another example is the joint channel estimation and passive beamforming framework in \cite{YZZ20}, where both are refined in each step with the CSI assumption for channel-gain-maximization. Therefore, in many scenarios, our perfect CSI assumption does not violate the applicability of our algorithms for future work in the literature.

\subsection{Performance Results with the Estimated CSI}
\begin{figure}[!t]
	\centering
	\includegraphics[width=0.45\textwidth]{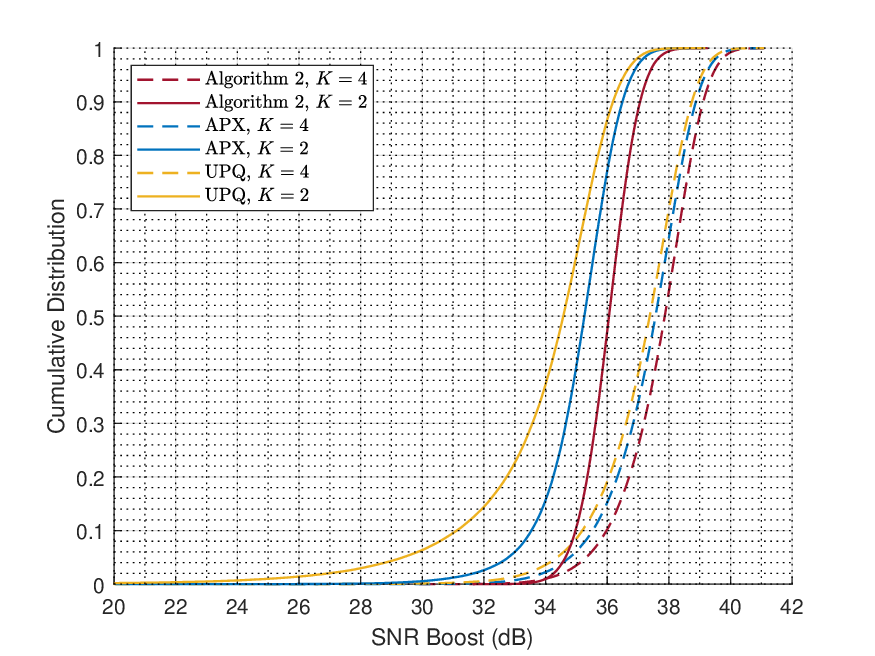}
	\caption{CDF plots for SNR Boost \cite{b1}, $N=200$, $\sigma^2_{\rm est\/} = -90{\rm dBm\/}$, $\kappa=0$, and $K \in \{2,4\}$.}
	\label{fig:CDFchest}
\end{figure}

We investigate the performance of our proposed algorithm also considering the effect of imperfect CSI, similar to \cite{ZSRLCL22}, by using the ON-OFF strategy in \cite{JD20}. In these results, the APX algorithm is from \cite{ZSRLCL22}, whereas UPQ and Algorithm~2 are as proposed in this paper. We adopt the same system parameters as in \cite{ZSRLCL22} to give a fair comparison, where the background noise power in the channel estimation phase $\sigma^2_{\text{est}}$ is set to -90 dBm as in the transmission phase. With this, the CDF of the performance results are given in Fig. \ref{fig:CDFchest} for $K=2$ and $K=4$ where the channels follow Rayleigh fading, i.e., $\kappa=0$. The performance results show that Algorithm~2 outperforms both UPQ and APX. The average gain against UPQ is 1.5 dB and against APX is 0.8 dB for $K=2$. When $K=4$, the average gain against UPQ is 0.5 dB and against APX is 0.3 dB. However, it is important to note that in the worst-case scenario, i.e., the lower SNR Boost regime, the gain of Algorithm~2 over UPQ and APX is much higher, especially when $K=2$.
\section{CONVERGENCE TO OPTIMALITY}
We will now discuss the convergence of Algorithm~2 and Algorithm~3 to the global optimum. This will be given in the {\em Theorem\/} below. The proof of our {\em Theorem\/} is similar to but actually different than the proof of {\em Theorem 1\/} of \cite{b1}.

{\em Theorem:\/} Algorithm~2 and Algorithm~3 yield the global optimum solution $(\theta_1^*,\theta_2^*,\ldots,\theta_N^*)$ to (\ref{eqn:eqn1}) in average time $\mathcal{O}(N)$.

{\em Proof:\/} First, we will discuss the global optimality of our Algorithm~2 and Algorithm~3. This follows because each $\theta_n$ is optimally decided from (\ref{eqn:lemma2}) as in {\em Proposition~1\/} and all the possible arcs have been considered for $\mu$ that yields to all unique values of $|g_c|$, for which the optimality follows from (\ref{eqn:eqn43}) and the text that follows it in Section~\ref{sec:nsteps}.

Next, we will discuss the complexity of Algorithm~2 and Algorithm~3. Similar to \cite{b1}, assuming $\alpha_n$ are uniformly distributed in $[0, 2\pi)$, $\varphi_n$ and $\varphi_m$ are uniformly distributed in the interval $[0, \frac{2\pi}{K})$. As in \cite{b1}, it can be argued that sorting in both algorithms will take $\mathcal{O}(N)$ time on average.
For Algorithm~2, the for loop from Step~6 to Step~13 takes $\sum_{l=1}^{N}\mathcal{O}(1) = \mathcal{O}(N)$ steps. For Algorithm~3, the for loop from Step~7 to Step~14 takes $\sum_{l=1}^{N-N'}\mathcal{O}(|{\cal R}_{m_l}|) = \mathcal{O}(N)$ steps. As our algorithms are working with elementwise updates, the time complexity of the steps is also linear in $N$, as discussed in Section \ref{sec:compcomp}. This means the overall complexity is $\mathcal{O}(N)$ on the average for both Algorithm~2 and Algorithm~3. Note that in contrast to \cite{b1}, when $h_0=0$, we do not need to try out all possible arcs, and the average complexity will remain as $\mathcal{O}(N)$, instead of becoming $\mathcal{O}(KN)$.\hfill$\blacksquare$

Although we intend Algorithm~1 as a conceptual stepping stone towards Algorithms~2 and~3, a proof for its convergence to the optimal solution can be deduced in a fashion similar to that for Algorithm~2.

Finally, we remark that our algorithms do not require the uniformity assumption on $\alpha_n$ to be able to assure the global optimum. We assume $\alpha_n$ to be uniform for calculating the complexity, as in \cite{b1}. In fact, both Algorithm~2 and Algorithm~3 can be applied to a scenario and assure the global optimum where $\alpha_n$ are arbitrarily selected. 
\section{COMPUTATIONAL TIME RESULTS} \label{sec:compcomp}
\begin{figure*}[!t]
	\centering
	\begin{minipage}{0.48\textwidth}
		\centering
		\includegraphics[width=0.8\textwidth]{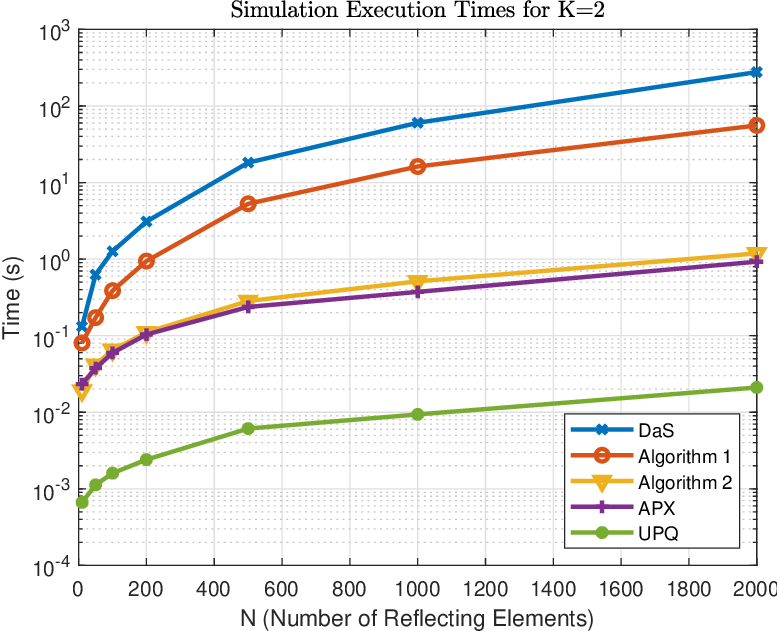}
		\caption{Plot of execution time [s] comparisons for 1000 channel realizations in NLOS, $K=2$.
			Note vertical scale is logarithmic.}
		\label{fig:figcck2}
	\end{minipage}%
	\hspace{0.03\textwidth}
	\begin{minipage}{0.48\textwidth}
		\centering
		\includegraphics[width=0.8\textwidth]{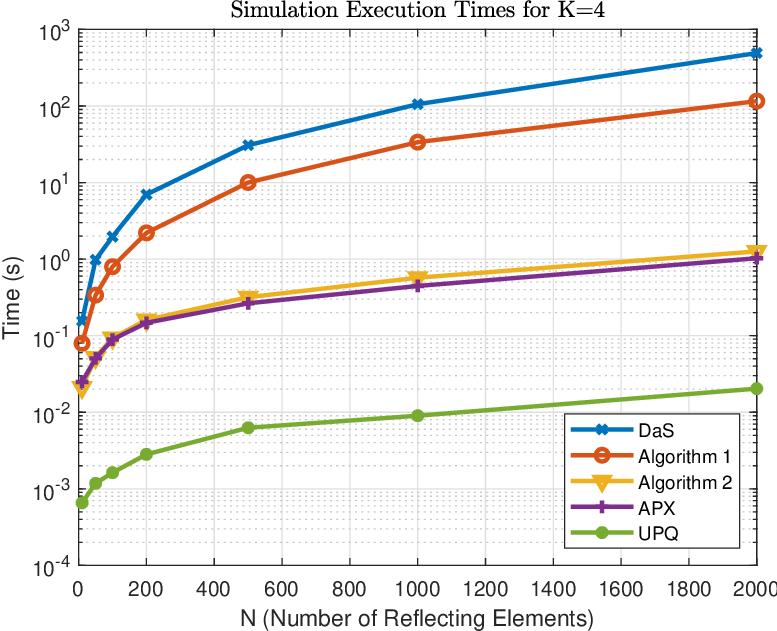}
		\caption{Plot of execution time [s] comparisons for 1000 channel realizations in NLOS, $K=4$.
			Note vertical scale is logarithmic.}
		\label{fig:figcck4}
	\end{minipage}
\end{figure*}

\begin{table*}[!h]
\begin{center}
\caption{Execution time [s] comparisons for 1000 channel realizations in non line-of-sight channel (NLOS), $K=2$.}
\label{tbl:cck2}
\begin{tabular}{lccccccc}
	\hline
	Method&$N=10$&$N=50$&$N=100$&$N=200$&$N=500$&$N=1000$&$N=2000$\\
	\hline
	\hline
	DaS&0.1319&0.6268&1.2641&3.0912&18.2086&60.3753&277.2977\\
	Algorithm~1&0.0804&0.1722&0.3874&0.9425&5.2790&16.2042&55.7506\\
	Algorithm~2&0.0192&0.0416&0.0650&0.1107&0.2836&0.5130&1.1875\\
	APX&0.0233&0.0377&0.0596&0.1036&0.2373&0.3732&0.9257\\
	UPQ&0.0007&0.0011&0.0016&0.0024&0.0061&0.0094&0.0211\\
	\hline
\end{tabular}
\end{center}
\begin{center}
\caption{Execution time [s] comparisons for 1000 channel realizations in NLOS, $K=4$.}
\label{tbl:cck4}
\begin{tabular}{lccccccc}
	\hline
	Method&$N=10$&$N=50$&$N=100$&$N=200$&$N=500$&$N=1000$&$N=2000$\\
	\hline
	\hline
	DaS&0.1552&0.9806&1.9666&6.9802&30.7800&105.6092&491.2088\\
	Algorithm~1&0.0796&0.3384&0.7967&2.2021&10.0270&33.7036&115.7444\\
	Algorithm~2&0.0211&0.0521&0.0934&0.1609&0.3180&0.5718&1.2623\\
	APX&0.0249&0.0506&0.0890&0.1475&0.2648&0.4460&1.0281\\
	UPQ&0.0007&0.0012&0.0016&0.0028&0.0063&0.0090&0.0204\\
	\hline
\end{tabular}
\end{center}
\end{table*}
We now provide computational complexity figures for our algorithms Algorithm~1 and Algorithm~2 in Fig.~\ref{fig:figcck2} and Fig.~\ref{fig:figcck4} for $K=2$ and $K=4$, respectively. We compare our algorithm with a number of algorithms from the literature, which are Approximation algorithm (APX) \cite{ZSRLCL22}, and Divide-and-Sort (DaS) algorithm \cite{XDMWQ22}. We also tabulate these results in terms of simulation time on the same computer
(Dell XPS 15 9530 employing Intel Core i9-13900H CPU, 2.6 GHz, with 14 cores and 20 logical processors) with implementations carried out in Matlab. We note that all the results provided are obtained with our own implementation of the algorithms in the most efficient way we were able to achieve. In Table~\ref{tbl:cck2}, we have the simulation time results in seconds plotted against the number of RIS elements $N$ for $K=2$. Then, in Table~\ref{tbl:cck4}, we have the simulation results for $K=4$. We note that among these algorithms, APX and UPQ do not have optimal performance, whereas DaS is claimed to be optimal. Recall that Algorithm~1 and Algorithm~2 achieve the optimal result.

In Fig.~\ref{fig:figcck2} and Fig.~\ref{fig:figcck4}, the computational complexities of the algorithms are plotted against the number of RIS elements, $N$. Algorithm~2, APX, and UPQ provides execution times that increase linearly with $N$. Moreover, it is seen that our Algorithm~2 can achieve comparable computational complexity with APX, for $K=2$ and $K=4$, thanks to our practical initialization technique presented in the Appendix. Therefore, Algorithm~2, or Algorithm~3 when applicable, can be considered as benchmark algorithms instead of the APX algorithm, to get the \textit{optimum} result with negligible additional computational costs.
DaS is claimed to be optimal, but requires extremely high execution time, which makes it difficult to increase $K$. Our algorithms Algorithm~1 and Algorithm~2 are optimal in terms of performance. With the substantial reduction in computational complexity from Algorithm~1, Algorithm~2 can achieve substantially lower execution times, which is robust to increasing $K$, as shown in Tables~\ref{tbl:cck2} and \ref{tbl:cck4}, still ensuring the optimum result.

\subsection{Remarks on the Complexity}
Finally, we note that, to achieve a linear time-complexity as in \cite{b1}, the main enabling factor is using the incremental update of $g_l$ with $\mathcal{N}(\lambda_l)$ in the for-loop of Algorithm 1, 2, and 3, rather than calculating $\theta_n, n=1,2,\dots,N$ for each candidate $\phase{\mu}$. In order to do the elementwise updates, the sorting of channels in the initialization steps of Algorithm~1, Algorithm~2, and Algorithm~3 are required. It could be possible to avoid sorting of the channels by quitting elementwise updates.

Assume we adopt an approach, similar to \cite{XDMWQ22},\cite{9961233}, where we first gather possible phase shift selections $\boldsymbol{\theta}^l, l=1,\dots,L$, and then perform each of them separately to find the maximum solution, for which the time complexity becomes $\mathrm{O}(N^2)$ \cite{b1}. With this approach, \cite{XDMWQ22},\cite{9961233} require the sorting algorithm, in fact, \cite{9961233} reports the complexity of the sorting requirement as it is the dominant term. However, we claim that, if we were to quit the elementwise approach, we could avoid the requirement of sorting in our algorithms. The main function of sorting is to try out all ``arcs'' for $\mu$, so, we need to know the sorting of $\phase{s_{nk}}$. On the other hand, if we were to try $\mu=s_{n^\prime k^\prime}$ for all unique $s_{n^\prime k^\prime}$ and decide $\theta_n, n\in \{1,\dots,N\} \setminus n^\prime$ with our \textit{Lemma}, then setting $\theta_{n^\prime}=(k^\prime-1)\omega$, we could generate the same set of solutions. Note that, in that case, the information of which $s_{nk}$ should come before or after would be unnecessary. Therefore, we could avoid sorting. However, it still would not perform better than Algorithm~1, as the elementwise update plays a significant role in terms of lower time complexity. Moreover, using our initialization technique with the least number of steps can reduce the overall computational time substantially with the elementwise updates.
\section{ALGORITHM COMPARISONS}
To highlight the novelty and contribution of our paper, we compare our work with the existing literature 
for the problem defined in (\ref{eqn:eqn1}) in Table~\ref{tab:comparison}. There are two main comparisons: The first is the number of search steps to ensure convergence to the global optimum, and the second is the required time complexity to run the algorithms. Note that, although the number of search steps is linear in $N$, an incremental or elementwise structure is required for trying all the candidate phase shift configurations. Among those algorithms with ensured global optimality, our Algorithm~2 and Algorithm~3 are the only ones that can ensure global optimality in $N$ or {\it fewer\/} search steps for any scenario. Regarding the algorithms that {\em converge to a local optimum,\/} APX requires three search steps to {\it approximate\/} the global optimum. Other algorithms that achieve a local minimum are CPP and our algorithm UPQ. The detailed comparison of CPP and UPQ is discussed in Sec.~\ref{ch:1}.

\begin{table}
	\centering
 	\caption{Comparison of Algorithms~2-3 and UPQ with algorithms from the literature. See the text for definition of ${\cal N}(\lambda_l)$.}
	\begin{tabular}{|c|c|c|c|}
                    \hline
					& Search Steps & Time  & Optimality\\
					&  				& Complexity & Guarantee\\
                    \hline\hline
	\cite{ZSRLCL22}	& 3 & $\mathcal{O}(N)$ & Local\\
	APX 			&  &  & \\
					\hline
	\cite{ZSRLCL22}	& Projection of Each & --- & Local\\
	CPP 			& Phase Selection &  & \\
                    \hline
                    	\cite{ZSRLCL22}	& $2N+2$, for $K=2$ & $\mathcal{O}(N)$ & Global\\
	Optimal			&  & ($K=2$ Only) & \\
                    \hline
	\cite{XDMWQ22}	& $N$, $K=2$ & $\mathcal{O}(N^2)$ &Global \\
			DaS		& $KN$, $K>2$ & (not elementwise) & \\
					\hline
	\cite{b1}		& $2N$, $h_0\neq0$ & $\mathcal{O}(N)$ & Global\\
					& $KN$, $h_0 = 0$ & $\mathcal{O}(KN)$ & \\
					\hline
    \cite{SBRFZT21}	& $N$, for $h_0 = 0$ & $\mathcal{O}(N^2)$ & Global\\
					&  & (not elementwise) & \\
					\hline
	\cite{9961233}	& $2N+1$, $K=2$ & $\mathcal{O}(N^2)$ & Global\\
					& $2N(K-1)$, $K>2$ & (not elementwise) & \\
					\hline
	\cite{LDDHCHB23}	& $KN$ & $\mathcal{O}(N^2)$ & Global\\
	FPB				&   & (not elementwise) & \\
					\hline
					\hline
	\textbf{UPQ}				& Deterministic & --- & Local\\
					&  &  & \\
					\hline
     \textbf{Algorithm}	& $\boldsymbol{N},$ {\em any\/} $h_0$ & $\boldsymbol{\mathcal{O}(N)}$ & Global\\
                    \textbf{2}&${\cal N}(\lambda_l)=1,$ all $l$ &  & \\
					\hline
    \textbf{Algorithm}	& $\boldsymbol{<N}$, {\em any\/} $h_0$ & $\boldsymbol{\mathcal{O}(N)}$ & Global\\
                    \textbf{3}&${\cal N}(\lambda_l) > 1,$ some $l$ &  & \\
					\hline
	\end{tabular}
\label{tab:comparison}
\end{table}

\section{CONCLUSION}
In this paper, we provided necessary and sufficient conditions for determination of optimum phase values in order to
maximize the received power at a UE which receives its transmission by means of reflections from an RIS, when the
phase values are from a discrete-valued set. Algorithms are provided to achieve this in a number of steps equal
to $N$, the number of RIS elements, or fewer. In the literatute, the number of steps to achieve this maximum is given
as $KN$ or $2N$ on the average, e.g., \cite{b1,XDMWQ22}. 
In conclusion, for a discrete-phase RIS, the techniques in this paper achieve the optimum received power in the
smallest number of steps published in the literature with an elementwise update rule. In addition, in each of those
$N$ steps, the techniques presented
determine only one or a small number of phase shifts, which result in a substantial reduction of computation
time, as compared to the algorithms in the literature, e.g., \cite{XDMWQ22,SBRFZT21,9961233}.

Finally, we want to make the following important point. In this paper, we addressed the ongoing problem
in the literature of finding an {\em optimal solution\/} to the problem (\ref{eqn:eqn1}) within the fewest
number of steps, or with minimum computational complexity. Our Algorithm~2 (or Algorithm~3 when
applicable) achieves this goal. Yet, a secondary result of this paper is that the intuitive UPQ solution,
which is based on the independent uniform quantization of $\theta_n$ on the unit circle with $K$
points, results in suboptimal but very close to the optimal solution with very small complexity for all
practical cases, provided full CSI.

\section{ACKNOWLEDGMENT}
The authors would like to thank Professor Thomas Ketseoglou for discussions on the
asymptotic analysis of the UPQ algorithm.
%
%
\section*{APPENDIX: ELEMENT-BASED SIMPLE UPDATE RULE}
We now further simplify Algorithm~1 by employing our periodicity proof, so that there is no need for calculating $s_{nk}$ or $\lambda_l .$
What we need to have is, given an initial $\phase{\mu}$ selection, say $e^{j \small \phase{\mu}\normalsize} \in {\rm arc}
(e^{j\lambda_{i-1}}:e^{j\lambda_i}),$ we want to know the $N$-step update rule ${\cal N}(\lambda_l),
l=i,i+1,\ldots,i+N-1$ in the for loop of Algorithm~1.

{\em Claim:\/} Let ${\cal U}$ be the set to define the $N$ consecutive updates in the for loop of Algorithm~1.
For an initial $\phase{\mu_0}$ selection, the update rule in the for-loop of Algorithm~1 will be ${\cal U} = \{ n_1,n_2,
\ldots, n_N | 0\le \varphi_{n_1} < \varphi_{n_2} < \cdots < \varphi_{n_N} < \frac{2\pi}{K}, \varphi_n =
(\alpha_n - \phase{\mu_0} + \frac{\pi}{K}) \;{\rm mod}\; \frac{2\pi}{K}, n = 1,2, \ldots, N\}$.

{\em Proof:\/}
First, consider the case when $\phase{\mu} = 0.$ We know that the initial arc is
${\rm arc}(e^{j\lambda_L}:e^{j\lambda_1})$. Therefore, the update rule must be
${\cal U} = ({\cal N}(\lambda_l))_{l=1}^N$. We have already calculated this in (\ref{eqn:eqn387}) for any
Case $i$ given in (\ref{eqn:eqn34}). Note, from (\ref{eqn:eqn34}) to (\ref{eqn:eqn387}),
$({\cal N}(\lambda_l))_{l=1}^{NK}$ follows from the indexes of the
sorted values of
\begin{equation}
\varphi_n = \left(\alpha_n+\frac{\pi}{K}\right) \;{\rm mod}\; \frac{2\pi}{K} .
\label{eqn:appb1}
\end{equation}
Now, consider the case when $\phase{\mu} = \phase{\mu_0}$ where $e^{j \small \phase{\mu_0} \normalsize} \notin {\rm arc}(e^{j\lambda_L}:e^{j\lambda_1})$. In this case,
instead of moving $\mu$ to a new arc, we can introduce an offset of $-\phase{\mu_0}$ for all $\lambda_l$.
Note that this corresponds to $\alpha_n\rightarrow \alpha_n - \phase{\mu_0},$ for all $n$. Therefore
(\ref{eqn:appb1}) will be updated as
\begin{equation}
\varphi_n = \left(\alpha_n - \phase{\mu_0} + \frac{\pi}{K}\right)\; {\rm mod}\; \frac{2\pi}{K}.
\label{eqn:appb2}
\end{equation}
Thus, the proof is complete.$\hfill\blacksquare$

Now, when $h_0 \neq 0$, to initialize with $\phase{\mu} = \alpha_0 - \frac{\pi}{K}$, we can simply insert $\phase{\mu_0}
= \alpha_0 -\frac{\pi}{K}$ in (\ref{eqn:appb2}) and get
\begin{equation}
\varphi_n = (\alpha_n - \alpha_0) \;{\rm mod}\; \frac{2\pi}{K}\label{eqn:eqn42}
\end{equation}
to be used in the initialization step. When the BS-UE link
is completely blocked, or $h_0=0$, initializations can be updated as $\phase{\mu}=0$ in Step~4 and
$\varphi_n = \left(\alpha_n-\frac{\pi}{K}\right)\; {\rm mod}\; \frac{2\pi}{K}$ for $n=1,2,\ldots,\frac{M}{K}$ in Step~2.

It is important to note that, the simplification in (\ref{eqn:appb2}) relieves Algorithm~1
from the burden to calculate $NK$ instances of both $s_{nk}$ and $\lambda_l$, and significantly reduces the computational complexity, as shown in Section~\ref{sec:compcomp}.

\bibliographystyle{IEEEtran}
\bibliography{ref}
\end{document}